\documentstyle[12pt]{article}   

\large
\newcommand{\beq}{\begin{equation}}
\newcommand{\eeq}{\end{equation}}

\makeatletter
\@addtoreset{equation}{section}
\makeatother

\newtheorem{Definition}{Definition}[section]

\def\be{\begin{equation}}
\def\ee{\end{equation}}
\def\ba{\begin{eqnarray}}
\def\ea{\end{eqnarray}}

\def\ag{{{\cal A}/{\cal G}}}

\def\agb{{\overline {{\cal A}/{\cal G}}}}

\def\a{\alpha}

\def\Comp{{\mathchoice
{\setbox0=\hbox{$\displaystyle\rm C$}\hbox{\hbox to0pt
{\kern0.4\wd0\vrule height0.9\ht0\hss}\box0}}
{\setbox0=\hbox{$\textstyle\rm C$}\hbox{\hbox to0pt
{\kern0.4\wd0\vrule height0.9\ht0\hss}\box0}}
{\setbox0=\hbox{$\scriptstyle\rm C$}\hbox{\hbox to0pt
{\kern0.4\wd0\vrule height0.9\ht0\hss}\box0}}
{\setbox0=\hbox{$\scriptscriptstyle\rm C$}\hbox{\hbox to0pt
{\kern0.4\wd0\vrule height0.9\ht0\hss}\box0}}}}
\def\Co{{\mathchoice
{\setbox0=\hbox{$\displaystyle\rm C$}\hbox{\hbox to0pt
{\kern0.4\wd0\vrule height0.9\ht0\hss}\box0}}
{\setbox0=\hbox{$\textstyle\rm C$}\hbox{\hbox to0pt
{\kern0.4\wd0\vrule height0.9\ht0\hss}\box0}}
{\setbox0=\hbox{$\scriptstyle\rm C$}\hbox{\hbox to0pt
{\kern0.4\wd0\vrule height0.9\ht0\hss}\box0}}
{\setbox0=\hbox{$\scriptscriptstyle\rm C$}\hbox{\hbox to0pt
{\kern0.4\wd0\vrule height0.9\ht0\hss}\box0}}}}
\def\Rl{{\mathchoice
{\setbox0=\hbox{$\displaystyle\rm R$}\hbox{\hbox to0pt
{\kern0.4\wd0\vrule height0.9\ht0\hss}\box0}}
{\setbox0=\hbox{$\textstyle\rm R$}\hbox{\hbox to0pt
{\kern0.4\wd0\vrule height0.9\ht0\hss}\box0}}
{\setbox0=\hbox{$\scriptstyle\rm R$}\hbox{\hbox to0pt
{\kern0.4\wd0\vrule height0.9\ht0\hss}\box0}}
{\setbox0=\hbox{$\scriptscriptstyle\rm R$}\hbox{\hbox to0pt
{\kern0.4\wd0\vrule height0.9\ht0\hss}\box0}}}}

\title{QSD IV :\\ 2+1 Euclidean Quantum Gravity as a model to test 
3+1 Lorentzian Quantum Gravity}
\author{T. Thiemann\thanks{thiemann@math.harvard.edu}
\thanks{New address :
Albert-Einstein-Institut, Max-Planck-Institut f\"ur Gravitationsphysik, 
Schlaatzweg 1, 14473 Potsdam, Germany, 
Internet : thiemann@aei-potsdam.mpg.de}\\
       Physics Department, Harvard University, \\
       Cambridge, MA 02138, USA}
\date{{\small \today\\ Preprint HUTMP-96/B-360}}

\begin{document}

\maketitle

\begin{abstract}
The quantization of Lorentzian or Euclidean 2+1 gravity by canonical methods
is a well-studied problem. However, the constraints of 2+1 
gravity are those of a topological field theory and therefore resemble very 
little those of the corresponding Lorentzian 3+1 constraints.

In this paper we canonically quantize Euclidean 2+1 gravity for arbitrary 
genus of the spacelike hypersurface with new, classically equivalent  
constraints that maximally probe the Lorentzian 3+1 situation. We choose 
the signature to be Euclidean because this 
implies that the gauge group is, as in the 3+1 case, $SU(2)$ rather than 
$SU(1,1)$. We employ, and carry out to full completion, the new quantization 
method introduced in preceding papers of this series which resulted in a 
{\em finite} 3+1 Lorentzian quantum field theory for gravity.

The space of solutions to all constraints turns out to be much larger 
than the one as obtained by traditional approaches, however, it is fully 
included. Thus, by suitable restriction of the 
solution space, we can recover all former results which gives confidence in 
the new quantization methods. The meaning of the remaining 
``spurious solutions" is discussed.
\end{abstract}

\section{Introduction}

The canonical quantization of 2+1 (pure) gravity is a well studied problem 
and 
the literature on this subject is extremely rich (see \cite{0} and 
references therein). It 
may appear therefore awkward to write yet another paper on this subject.

The point of this paper is to quantize 2+1 gravity by starting with a 
new Hamiltonian (constraint) rather than the one that imposes 
flatness of the connection (see, for instance, \cite{1}). Therefore, we 
are actually dealing with a new field theory. The reason why we still 
can call this theory 2+1 gravity (although in the Euclidean signature) is 
because classically both theories are equivalent, it is in the quantum 
theory only where discrepancies arise.

The motivation to study this model comes from 3+1 gravity : In \cite{2} a 
new method is introduced to quantize the Wheeler-DeWitt constraint for 
3+1 Lorentzian gravity and one arrives at a {\em finite} quantum field 
theory. It is therefore of interest to check whether that quantum theory
describes a physically interesting phase of the full theory of quantum 
gravity. One way to do that is to apply the formalism to a model system which
maximally tests the 3+1 theory while being completely solvable.

It is often said that 2+1 gravity in its usual treatment as for instance in
\cite{1} is such a model which tests the 3+1 theory in various technical 
and conceptual ways. The author disagrees with such statements for 
a simple reason :\\
The constraints of usual 2+1 gravity and of 3+1 gravity are not even 
algebraically similar. Thus, one has to expect that the resulting 
quantum  theories are mutually singular in a certain sense. We will find 
that this expectation turns out to be correct.\\
One can partially fix this by studying {\em Euclidean} 2+1 gravity to
test {\em Lorentzian} 3+1 gravity because then the two gauge groups 
($SU(2)$) coincide, in the Lorentzian signature the gauge group of 
2+1 gravity would be $SU(1,1)$. However, 
this is not enough : while now the Gauss 
constraints of both theories generate the same motions the rest of the 
constraints are still very different with respect to each other. More 
precisely, the 2+1 remaining constraint says that the connection $A$ is 
flat, 
that is, its curvature $F$ vanishes. Thus it does not involve the momenta $E$
conjugate to the connection {\em at all}. The situation in the 3+1 theory 
is very different : here we have as the remaining constraints a 
constraint that generates diffeomorphisms and the famous Wheeler-DeWitt
constraint. Both constraints depend on the momenta, the Wheeler-DeWitt 
constraint even non-analytically. In \cite{3} the authors propose to 
quantize the constraints $FE=FEE=0$. However, this has never been done 
in the literature, one reason being that the $FEE$ constraint is as 
difficult to quantize as in the the 3+1 case. Moreover, the two constraints
$FE=FEE=0$ are equivalent to the $F=0$ constraint only when the 
two-metric $q$ 
is non-singular, that is, $\det(q)>0$ and therefore it is no surprise that 
the two 
theories are not even classically equivalent as was shown in \cite{4} (for
the theory defined by the $F=0$ constraint the condition $\det(q)>0$ is put
in by hand in order to have Euclidean signature). 

In this paper we are using the constraints $FE=FEE/\sqrt{\det(q)}=0$. 
There are several reasons that speak for this choice :\\
First of all these 
constraints are at least classically completely equivalent to the 
$F=0$ constraints because clearly they make sense only when $\det(q)>0$.
In fact we will show that there is a field dependent non-singular map
between the Lagrange multipliers of the two theories which map the two 
sets of constraints into each other.\\
Secondly, they are just as in the 3+1 theory non-analytic in $E$ (because
$\det(q)$ is a function of $E$) and so will test this feature of the 
3+1 theory as well. In particular, both constraints are densities of weight
one and only constraints of this type have a chance to result in 
densely-defined diffeomorphism covariant operators as argued in \cite{2}.\\
Thirdly, these constraints are maximally in analogy to all the 3+1 
constraints.\\
\\
The plan of the present paper is as follows :

In section 2 we review the classical theory of Euclidean 2+1 gravity and
outline our main strategy of how to arrive at a well-defined Hamiltonian
constraint operator.

In section 3 we review the necessary background information on the 
the mathematical tools that have been developed for diffeomorphism 
invariant theories of connections. Those Hilbert space techniques 
are identical for the 2+1 and 3+1 theory so that we have one more reason 
to say that the model under consideration tests the 3+1 situation.
Also we need to construct a volume operator which as in the 3+1 theory 
plays a key role in the regularization of the (analog of the) Wheeler-DeWitt
constraint operator. The 2+1 volume operator turns out to be much less 
singular than the 3+1 operator which has some important impact on the
regularization of the constraint operators.

In section 4 we regularize the Wheeler-DeWitt operator. Many of the details
are exactly as in the 3+1 theory although there are some crucial 
differences coming from the lower dimensionality of spacetime and also
from the different singularity structure of the volume operator.

In section 5 we perform various consistency checks on the 2+1 Wheeler-DeWitt
operator obtained, in particular whether it is a linear, covariant and 
anomaly-free operator.

In section 6 we construct the full set of solutions to all constraints.
It his where we encounter, besides reassuring results that give faith 
in the programme started in \cite{2}, several surprises : 
\begin{itemize}
\item The quantum theory admits solutions which correspond to degenerate 
metrics. This happens although classically such solutions do not exist
given our constraints. This should not be confused with the situation in
\cite{4} because there degenerate metrics are allowed even at the 
classical level.
\item We find an uncountable number of rigorous distributional solutions to 
all constraints
which reveal an uncountable number of quantum degrees of freedom just as 
in any field theory with local degrees of freedom. This is in complete 
contrast to the usual treatment via the $F=0$ constraints which results 
in a topological quantum field theory with only a finite number of degrees
of freedom.
\item The space of solutions contains the solutions to the quantum $F=0$
constraints as a tiny subspace. 
This subspace of solutions can be equipped with an inner product
which is precisely the one that one obtains in traditional approaches.
This is reassuring that our methods lead
to well-established results and do not describe some unphysical phase of 
the theory. 
\item The huge rest of the solutions cannot be equipped with the inner 
product appropriate for the $F=0$ constraints because they do not 
correspond to measurable functions with respect to the corresponding 
measure. However, there is another natural
inner product available with respect to which they are normalizable. 
This inner product is likely to be the one that is appropriate also 
for the physically interesting solutions of the 3+1 constraints. The
solutions to the $F=0$ constraint in turn are not normalizable with 
respect to this second inner product. Thus as expected, the two sets of 
constraints have solution spaces which lie in the same space of 
distributions but they cannot be given the same Hilbert space topology. It 
is in this sense that the quantum theories are mutually singular.
\end{itemize}

In section 7 we conclude with some speculations 
of what the present paper teaches us for the 3+1 theory with regard to
the solutions that are spurious from the point of view of the $F=0$ 
constraint.

In the appendix we compute the spectrum of the 2+1 volume operator for 
the simplest states.\\
\\
Throughout the paper we mean by the wording ``2+1 or two-dimensional" always
2+1 Euclidean gravity while by ``3+1 or three-dimensional" we always mean 
3+1 Lorentzian gravity.

\section{Classical Theory}

Let us start by reviewing the notation (see, for instance, \cite{1}).\\
We assume that the three-dimensional spacetime is of the form 
$M=\Rl\times\Sigma$ where $\Sigma$ is a two-dimensional manifold of 
arbitrary topology, for instance, a compact, connected two-dimensional 
smooth manifold, that is, a Riemann surface of genus $g$ or an 
asymptotically flat manifold. Let $e_a^i$
be the co-dyad on $\Sigma$ where $a,b,c,..=1,2$ denote tensor indices and
$i,j,k,..=1,2,3$ denote $su(2)$ indices. The fact that we are dealing with
$su(2)$ rather than $su(1,1)$ implies that the two-metric $q_{ab}:=
e_a^i e_b^i$ has Euclidean signature. Moreover, let $A_a^i$ be an $su(2)$
connection and define the field $E^a_i:=\epsilon^{ab}e_b^i$ where 
$\epsilon_{ab}$ is the metric-independent totally skew tensor of density 
weight $-1$. Then it 
turns out that the pair $(A_a^i,E^a_i)$ is a canonical one for the 
Hamiltonian formulation of 2+1 gravity based on the Einstein Hilbert action
$S=\int_M d^3x\sqrt{|\det(g)|} R^{(3)}$ where $g$ is the three-metric and 
$R^{(3)}$ its scalar curvature.
In other words, $E^a_i$ is the momentum conjugate to $A_a^i$ so that the 
symplectic structure is given by
\be \label{1}
\{A_a^i(x),E^b_j(y)\}=\delta_a^b\delta^i_j\delta(x,y)\;.
\ee
The Hamiltonian of the theory is a linear combination of constraints,
$\int d^2x (\Lambda^i G_i+N^i C_i)$ for some Lagrange multipiers 
$\Lambda^i,N^i$ where
\ba \label{2}
G_i&:=&D_a E^a_i=\partial_a E^a_i+\epsilon_{ijk}A_a^j E^a_k
\mbox{ : Gauss constraint,}\nonumber\\
C_i&:=& \frac{1}{2}\epsilon^{ab}F_{ab}^i \mbox{ : Curvature constraint}
\ea
where $F_{ab}$ denotes the curvature of $A_a$. The Gauss constraint 
appears also in 3+1 gravity, however, the curvature constraint is 
completely different from the constraints that govern 3+1 gravity, \cite{2}.
The equivalent of $C_i$ in 3+1 gravity are two types of constraints, one 
of them, $V_a$, generates diffeomorphisms, the other one, $H$, generates
dynamics. The curvature constraints $C_i$ on the other hand do not generate
any such gauge transformations, in fact, the connection Poisson-commutes with
$C_i$ and shows that it is a Dirac observable with respect to $C_i$. The 
constraint $C_i=0$ imposes that the connection should be flat and thus 
the classically reduced phase space becomes the cotangent bundle over the
moduli space of flat $su(2)$ connections which is finite-dimensional.\\
It is obvious that the quantization of the model as defined by (\ref{2})
will not give too much insight into the 3+1 situation. In the following 
we will reformulate (\ref{2}) in such a way that it brings us in 
connection with 3+1 gravity.\\
It will turn out that the following compound field, called the 
{\em degeneracy vector}, for reasons that will become obvious soon
\be \label{3}
E^i:=\frac{1}{2}\epsilon^{ijk}\epsilon_{ab} E^a_j E^b_k
\ee
is a crucial one. Let us compute the square of this density of weight one :
\ba \label{4}
E^i E^i &=&\frac{1}{2}\epsilon_{ab}\epsilon_{cd}[E^a_i E^c_i][E^b_j E^d_j]
\nonumber\\
&=&\frac{1}{2}q_{bd}[E^b_j E^d_j]=\frac{1}{2}q_{bd} q_{ac}
\epsilon^{ba}\epsilon^{dc}
\nonumber\\
&=& \det(q),
\ea
that is, the two-metric is degenerate if and only if $E_i=0$ is identically
zero. We also see that $\det(q)$ is manifestly non-negative. Notice that
$E^a_i E^i=0$.\\
Whenever the degeneracy vector is non-vanishing we can perform the following
non-singular transformation $(N^i)\leftrightarrow(N^a,N)$ for a vector field
$N^a$, called the shift, and a scalar function $N$, called the lapse :
\be \label{5}
N^i=N^a \epsilon_{ab} E^b_i+N\frac{E^i}{\sqrt{\det(q)}} 
\Leftrightarrow
N^a=\epsilon_{ijk}\frac{E^i E^a_j}{\det(q)} N^k,\; 
N=\frac{N^i E^i}{\sqrt{\det(q)}}\;.
\ee
Notice that formula (\ref{5}) respects that $N^i,N^a,N$ have density weight
zero. Using (\ref{5}), we can now write the curvature constraint in the form
\ba \label{6}
N^i C_i &=& N^a V_a+N H \mbox{ where}\nonumber\\
V_a &:=& F_{ab}^i E^b_i\mbox{ : Diffeormorphism constraint}\nonumber\\
H &:=& \frac{1}{2} F_{ab}^i\epsilon_{ijk}\frac{E^a_j E^b_k}{\sqrt{\det(q)}}
\mbox{ : Hamiltonian constraint}\;.
\ea
Apart from the fact that we are in two rather than three space dimensions
these are precisely the constraints of Euclidean 3+1 gravity \cite{2}. 
Since the 3+1 Euclidean Hamiltonian constraint operator plays a {\em key 
role} in the quantization of the 3+1 Lorentzian Hamiltonian constraint
\footnote{In fact, once one has densely defined the Euclidean operator in 3+1
dimensions the fact that the Lorentzian operator in 3+1 is densely defined is a
simple Corollary \cite{2}.},
we claim that the set of constraints (\ref{6}) bring us in maximal 
contact with the 3+1 theory.\\
Notice that unlike in \cite{3}, we have a factor of $1/\sqrt{\det(q)}$
in the definition of the Hamiltonian constraint. This difference has two 
important consequences : 
\begin{itemize}
\item[1)] Classical :\\
The denominator in $H=F_i E^i/\sqrt{\det(q)}$ where $F_i:=\epsilon^{ab}
F_{ab}^i/2$ (or $F_{ab}^i=\epsilon_{ab} F^i$) blows up as $E^i$ vanishes. 
Since the limit $\lim_{\vec{E}\to 0}\vec{E}/||\vec{E}||$ depends on the 
details of the limiting procedure we {\em must exclude} degenerate metrics
{\em classically}. This is in contrast to \cite{4} where the authors exploit 
the possible classical degeneracy of the metric when one discards the 
denominator to demonstrate that one has already an infinite number 
of degrees of freedom at the classical level (notice, however, that their 
solutions, where $F^i$ or 
$E^i$ become null, do not apply since we are dealing with $su(2)$). 
\item[2)] Quantum :\\
It is by now known that one of the reasons for why 
$\tilde{H}:=\sqrt{\det(q)}H$ suffers from huge problems upon quantizing it
is due to the fact that $\tilde{H}$ has density weight two rather than one.
As argued in \cite{2}, only densities of weight one have a chance to 
be promoted into densely defined, covariant operators. This is why we 
{\em must} keep the denominator $1/\sqrt{\det(q)}$ in (\ref{6}) {\em
at the quantum level}. 
\end{itemize}
Just like in 3+1 gravity we wish to work in a connection representation, 
that is, states are going to be functions of connections. Then an immediate
problem with $H$ is that one has to give a meaning to the denominator
$1/\sqrt{\det(q)}$. In \cite{2} that was achieved for Lorentzian 
3+1 gravity by noting that the 
denominator could be absorbed into a Poisson bracket with respect to a 
functional $V$ of $q_{ab}$. The idea was then 
to use the quantization rule that Poisson brackets should be replaced by
commutators times $1/(i\hbar)$ and to replace $V$ by an appropriate operator
$\hat{V}$. Such an operator indeed exists and it is densely defined.

Is a similar trick also available for 2+1 gravity ? At first sight the 
answer seems to be in the negative because the underlying reason for why
such a trick worked for 3+1 gravity was that the co-triad $e_a^i$, the 
precise analogue of the degeneracy vector $E^i$,  considered 
as a function of $E^a_i$ was integrable, the generating functional being 
given by the total volume $V$ of $\Sigma$. In other words, we had
\be \label{7}
e_a^i=\frac{1}{2}\epsilon^{ijk}\epsilon_{abc}\frac{E^b_j 
E^c_k}{\sqrt{\det(q)}}=\frac{\delta V}{\delta E^a_i}\mbox{ with }
V:=\int_\Sigma d^3 x \sqrt{\det(q)}\;
\ee
However, if we take over the definition of $V$ (with $d^3x$ replaced by 
$d^2x$) then we find instead
\be \label{8}
\{A_a^i,V\}=\frac{\delta V}{\delta E^a_i}=\frac{q_{ab}E^b_i}{\sqrt{\det(q)}}
\mbox{ with } V:=\int_\Sigma d^2x \sqrt{\det(q)}\;.
\ee
Thus, there seems not such a trick available in the 2+1 case. However, it is
a matter of straightforward computation to verify that indeed
\be \label{9}
E^i=\frac{1}{2}\epsilon^{ab}\epsilon_{ijk}\{A_a^j,V\}\{A_b^k,V\}
\ee
which does not seem to help much because what we need is
$E^i/\sqrt{\det(q)}$ rather than $E^i$ itself. \\
The new input needed here as compared to the 3+1 case is as follows :
Notice that if we could replace $\sqrt{\det(q)}$ by V then we could absorb
it into the Poisson brackets by using the identity
$$
\frac{\{A_a^j,V\}\{A_b^k,V\}}{V}=4\{A_a^j,\sqrt{V}\}\{A_b^k,\sqrt{V}\}
$$
As we will see, $V$ can be promoted, just as in the 3+1 case, into a 
densely defined {\em positive semi-definite} operator. Therefore its
square root exists and it would follow that the last equation with 
Poisson brackets replaced by commutators would make sense as an operator.
In the next section we will define a Hilbert space and the corresponding
operator.\\
What remains is to justify the replacement of $\sqrt{\det(q)}$ by
$V$. That this is possible we will show in the section after the following.
It happens because the Poisson bracket gives a local quantity and therefore
we may actually replace $V$ by $V(x,\epsilon)$ in
$$\{A_a^i(x),V\}\equiv\{A_a^i(x),V(x,\epsilon)\}
\mbox{ where }V(B)=\int_B d^2x \sqrt{\det(q)}
$$
is the volume of a compact region $B$ and $V(x,\epsilon)$ is the volume 
of an 
arbitrarily small open neighbourhood of the point $x$, the smallness governed
by $\epsilon$. It is then easy to see that $\lim_{\epsilon\to 0}
V(x,\epsilon)/\epsilon^2=\sqrt{\det(q)}(x)$. Now, in the quantum theory 
we are going to point split the quantity $H$ and we will use a regularized
$\delta$ distribution with point split parameter $\epsilon$. As we will
see, that parameter can be absorbed into $V(x,\epsilon)$ to serve as a 
replacement for $\sqrt{(\det(q)}(x)$. 
The details are displayed in the 
following sections.

\section{Quantum Theory and Volume operator}

In this section we will review the definition of a Hilbert space for
diffeomorphism invariant theories of connections \cite{5}. This will be
our kinematical framework. On that Hilbert space we are going to construct
a 2+1 volume operator which turns out to be actually more complicated 
than the one for the 3+1 theory \cite{6,7}.

\subsection{Quantum kinematics}

In what follows we give an extract from \cite{5,5a}. The reader 
interested in the details is urged to study those papers.

We will denote by $\gamma$ a finite piecewise analytic graph in $\Sigma$. 
That is, we have analytic edges $e$ which are joined in vertices $v$.
We subdivide each edge into two parts and equip each part with an orientation
that is outgoing from the vertex (the point where these two parts meet is
a point of analyticity and therefore not a vertex of a graph, thus each 
edge from now on can be viewed to be incident at precisely one vertex).
Given an $su(2)$ connection $A_a^i$ on $\Sigma$ we can compute its holonomy
(or path-ordered exponential) $h_e(A)$ along an edge $e$ of the graph.
Recall that all representations of $SU(2)$ are completely reducible and 
that the (equivalence class of equivalent) irreducible ones can be 
characterized by a half integral non-negative number $j$, the spin of 
the representation. We will denote the matrix elements of the 
$j$-representation at $g\in SU(2)$ by $\pi_j(g)$.

Consider now a vertex $v$ of the graph and the edges $e_1,..,e_n$ incident
at $v$, that is, the graph has valence $n$. Under a gauge transformation 
$g$ at $v$ the holonomy transforms as
$h_{e_i}\to gh_{e_i},\;i=1,..,n$. Now consider the transformation of
the following function 
$$
\otimes_{i=1}^n \pi_{j_i}(h_{e_i})\to
[\otimes_{i=1}^n \pi_{j_i}(g)]\cdot
[\otimes_{i=1}^n \pi_{j_i}(h_{e_i})]\;.
$$ 
We are interested in making this function gauge invariant at $v$. To that end
we orthogonally decompose the tensor product of the $\pi_{j_i}(g)$ into 
irreducibles 
and look for the independent singlets in that decomposition. There is an 
orthogonal projector $c_v$ on each of these singlets, we say that it is 
compatible with the spins $j_1,..,j_n$, and 
so we can make our function gauge invariant at $v$ by contracting :
$c_v\cdot\otimes_{i=1}^n \pi_{j_i}(h_{e_i})$.

If we do that for each 
vertex we obtain a completely gauge invariant function called a
{\em spin-network function}. Thus a spin-network function is labelled by 
a graph $\gamma$, a colouring of its edges $e$ with a spin $j_e$ and 
a dressing of each vertex $v$ with a gauge-invariant projector $c_v$.
If we denote by $E(\gamma),V(\gamma)$ the set of edges and vertices of 
$\gamma$ respectively then we use the shorthand notation
$$
T_{\gamma,\vec{j},\vec{c}}
\mbox{ where }
\vec{j}:=\{j_e\}_{e\in E(\gamma)},\;
\vec{c}:=\{c_v\}_{v\in V(\gamma)},\;
$$ 
for that spin-network function.

The Hilbert space ${\cal H}$ that we are going to use for gauge invariant 
functions 
of connections is most easily described by saying that the set of all 
spin-network functions is a complete orthonormal basis of $\cal H$ (so
each spin-network function comes with a specific finite normalization 
factor). Notice that therefore $\cal H$ is not separable. Another 
characterization of $\cal H$ which is very useful is to display it as a 
certain $L_2$ space. To that end, consider the finite linear combinations 
$\Phi$ of spin-network functions. $\Phi$ can be turned into an Abelian
$C^\star$ algebra by saying that involution is just complex conjugation 
and by completing it with respect to the $sup-$norm over the space 
$\ag$ of {\em smooth} connections modulo gauge transformations. That 
$C^\star$ algebra is isometric 
isomorphic by standard Gel'fand techniques to the $C^\star$ algebra of 
continuous functions $C(\agb)$ where $\agb$ is the set of all 
homomorphisms from the original algebra into the complex numbers. The space
$\agb$, as the notation suggests, is a certain extension of $\ag$ and 
will be called the set of {\em distributional} connections. Indeed, it is 
the maximal extension such that (the Gel'fand transform of the) 
spin-network functions
are continuous. By standard results, the resulting topology is such that
$\agb$ is a compact Hausdorff space and as such positive linear 
functionals $\Gamma$ on $C(\agb)$ are in one to one correspondence with
regular Borel measures $\mu$ on $\agb$ via $\Gamma(f)=:\mu(f)=\int_\agb 
d\mu f$.\\
Now the measure $\mu_0$ underlying $\cal H$ is completely characterized 
by the integral of spin-network functions and is given by
$\mu_0(T_{\gamma,\vec{j},\vec{c}})=1$ if $T_{\gamma,\vec{j},\vec{c}}=1$ 
and $0$ otherwise. So we have ${\cal H}=L_2(\agb,d\mu_0)$ and 
spin-network functions play the same role for $\mu_0$ that Hermite 
functions play for Gaussian measures.

In the sequel we will topologize the space $\Phi$ of finite linear 
combinations of spin-network functions in a different way and we will call
$\Phi$ henceforth the space of cylindrical functions. A function 
$f_\gamma$ is said 
to be cylindrical with respect to a graph $\gamma$ if it can be written as
a finite linear combination of spin-network functions on that $\gamma$. The 
norm of $f_\gamma$ will be the $L_1$ norm $||f_\gamma||_1=\sum_I
|<T_I,f>|$ which equips $\Phi$ with the structure of a topological vector 
space. The distributional
dual $\Phi'$ is the set of all continuous linear functionals on 
$\Phi$. 
Certainly every element of $\cal H$ is an element of $\Phi'$ by the Schwarz
inequality and every element of $\Phi$ is trivially an element of $\cal H$.
Thus we have the inclusion $\Phi\subset{\cal H}\subset\Phi'$ (this is not 
a Gel'fand triple in the strict sense because the topology on $\Phi$ is 
not nuclear).
 
This furnishes the quantum kinematics. Notice that we can take over the 
results 
from \cite{5} without change concerning the Diffeomorphism constraint :
given an analyticity preserving diffeomorphism $\varphi$ we have a unitary
operator on $\cal H$ which acts on a function cylindrical with respect to 
a graph as $\hat{U}(\varphi)f_\gamma=f_{\varphi(\gamma)}$, that is, the 
diffeomorphism group $\mbox{Diff}(\Sigma)$ is unitarily represented.
This implies that one can group average with respect to the 
diffeomorphism group as in \cite{5}. We will return to this point in section
6.

\subsection{The 2+1 volume operator}

The plan of this subsection is as follows :\\
Since $\hat{E}^i(x)$ is a density of weight one, it makes sense that it
will give rise to a well-defined and diffeomorphism-covariantly defined
operator valued distribution. In a second step we will point-split
$\det(q)=E^i E^i$ and take the square root of the resulting 
operator. Again, since $\sqrt{\det(q)}$ is a density of weight one,
it can be turned into a well-defined operator-valued distribution even in 
regulated form and the limit as the regulator is removed exists.\\
Let us then begin with $E^i$. Let as in the previous section $f_\gamma$
denote a function cylindrical with respect to a graph $\gamma$ and denote
by $E(\gamma)$ its set of edges. Edges are, by suitably subdividing them 
into two halves, in the sequel always supposed to be oriented as outgoing 
at a vertex. We will compute the action of various operators first on 
functions of smooth connections and then extend the end result to all of 
$\agb$.\\ 
Let $\delta_{\vec{\epsilon}}(x,y)=
\delta_{\epsilon_1}(x^1,y^1)
\delta_{\epsilon_2}(x^2,y^2)$ be any 
two-parameter family of smooth functions of compact support such that 
$\lim_{\epsilon_1,\epsilon_2\to 0}\int_\Sigma 
d^2y \delta_\epsilon(x,y)f(y)=f(x)$
for any, say smooth, function on $\Sigma$ where $\vec{\epsilon}=
(\epsilon_1,\epsilon_2)$ parametrizes 
the size of the support. Consider the point-split operator 
\be \label{21}
\hat{E}^i_{\vec{\epsilon},\vec{\epsilon}'}(x)
:=\frac{1}{2}\epsilon_{ab}\epsilon^{ijk}\int_\Sigma d^2y\int_\Sigma d^2z 
\delta_{\vec{\epsilon}}(x,y)\delta_{\vec{\epsilon}'}(x,z)
\hat{E}^a_j(y)\hat{E}^b_k(z)
\ee
and apply it to $f_\gamma$. Notice that upon replacing 
$\hat{E}^a_i(x)=-i\hbar\delta/\delta A_a^i(x)$
\be \label{22}
\hat{E}^a_i(x)f_\gamma=-i\hbar\sum_{e\in E(\gamma)}\int_0^1 dt \delta(x,e(t))
\dot{e}^a(t)X^i_e(t) f_\gamma
\ee
where $X^i_e(t)=\mbox{tr}([h_e(0,t)\tau_i h_e(t,1)]^T\partial/\partial 
h_e(0,1))$, $h_e(a,b)$ is the holonomy from parameter value $a$ to $b$
and $\tau_i$ are generators of $su(2)$ with structure constants 
$\epsilon_{ijk}$. We also need the quantity\\
$X^{ij}_e(s,t)=\mbox{tr}([h_e(0,s)\tau_i h_e(s,t)\tau_j h_e(t,1)]^T
\partial/\partial h_e(0,1))$ for $s<t$ (modulo $1$). Then it is easy to 
see that 
\ba \label{23}
&&\hat{E}^i_{\vec{\epsilon},\vec{\epsilon}'}(x)f_\gamma\nonumber\\
&=&-\hbar^2\frac{1}{2}\epsilon_{ab}\epsilon^{ijk}
\sum_{e,e'\in E(\gamma)}\int_\Sigma 
d^2y\int_\Sigma d^2z \delta_{\vec{\epsilon}}(x,y)
\delta_{\vec{\epsilon}'}(x,z)\times
\nonumber\\
&\times& \int_0^1 dt \delta(y,e(t)) \dot{e}^a(t)
\int_0^1 dt' \delta(z,e'(t')) \dot{e}^{\prime b}(t')
\times\nonumber\\
&\times& [X^k_{e'}(t') X^j_e(t)+\delta_{e,e'}\{
\theta(t,t')X^{jk}_e(t,t')+\theta(t',t)X^{jk}_e(t',t)\}] f_\gamma
\nonumber\\
&=&-\hbar^2\frac{1}{2}\epsilon_{ab}\epsilon^{ijk} \sum_{e,e'\in E(\gamma)}
\int_0^1 dt  \dot{e}^a(t)
\int_0^1 dt' \dot{e}^{\prime b}(t')
\delta_{\vec{\epsilon}}(x,e(t))\delta_{\vec{\epsilon}'}(x,e'(t'))\times
\nonumber\\
&\times& [X^k_{e'}(t') X^j_e(t)+\delta_{e,e'}\{
\theta(t,t')X^{jk}_e(t,t')+\theta(t',t)X^{jk}_e(t',t)\}] f_\gamma
\ea
where $\theta(s,t)=1$ if $s<t$ and $0$ otherwise.\\ 
We are now interested 
in the limit $\epsilon\to 0$ and proceed similar as in \cite{7}. We must
adapt the regularization to each pair $e,e'$ to get a well-defined result.
\begin{itemize} 
\item[1)] Case $e=e'$.\\ 
If $x$ does not lie on $e$ then for sufficiently small $\vec{\epsilon}$ we 
must get 
$\delta_{\vec{\epsilon}}(x,e(t))=0$ for any $t\in [0,1]$. Thus in the 
limit we get a 
non-vanishing contribution if and only if there exists a value $t_x\in [0,1]$
such that $e(t_x)=x$ (there is at most one such value $t_x$ because edges 
are not self-intersecting). Since $\dot{e}$ is nowhere vanishing we must
have $\dot{e}^1(t_x)\not=0$ (switch $1\leftrightarrow 2$ if necessary).
We send $\epsilon_1,\epsilon_1'\to 0$ and find that
$\delta_{\vec{\epsilon}}(x,e(t))\to \delta_{\epsilon_2}(x^2,e^2(t))
\delta(t-t_x)/|\dot{e}^1(t_x)|$ and similar for 
$\delta_{\vec{\epsilon}'}(x,e'(t'))$. Inserting this into (\ref{23})
we find that there is no contribution for $e=e'$ because of the two zeroes
$0=\epsilon_{ab}\dot{e}^a(t_x)\dot{e}^b(t_x)$ and
$0=\epsilon_{ijk}[X^{ij}_e(t_x,t_x)+X^{ji}_e(t_x,t_x)]$. Notice that it was 
crucial to have $\epsilon_2,\epsilon_2'$ still finite as otherwise the 
appearing $\delta_{\epsilon_2}(0)\delta_{\epsilon_2'}(0)$ would be 
meaningless.
\item[2)] Case $e\not=e'$. \\
If again $x$ does not lie on both $e,e'$ then by choosing $\vec{\epsilon},
\vec{\epsilon}'$ sufficiently small we must get zero. Therefore $e,e'$
must intersect and as we have divided edges into two halves they can 
intersect at most in their common starting point corresponding to $t=t'=0$ 
which is thus a vertex $v$ of the graph $\gamma$. 
\begin{itemize}
\item[A)] Subcase\\
Consider first the case 
that $e,e'$ have co-linear tangents at $t=0$ and let us assume that
$\dot{e}^1(0),\dot{e}^{\prime 1}(0)\not=0$ (switch $1\leftrightarrow 2$ 
if necessary). Then we first send $\epsilon_1,\epsilon_1'\to 0$ which 
results in 
$$
\delta_{\vec{\epsilon}}(x,e(t))
\delta_{\vec{\epsilon}'}(x,e'(t'))
\to 
\delta_{\epsilon_2}(x^2,e^2(t))
\delta_{\epsilon_2'}(x^2,e^{\prime 2}(t'))
\frac{\delta(t)\delta(t')}{|
\dot{e}^1(0)\dot{e}^{\prime 1}(0)|}
$$
and thus performing the two $t$ integrals we get zero as above because
$0=\epsilon_{ab}\dot{e}^a(0)\dot{e}^{\prime b}(0)$ by assumption.
\item[B)] Subcase\\
We are left with the case that the tangents of $e,e'$ are linearly 
independent at $x=v$. We replace 
$\delta_{\vec{\epsilon}}(x,e(t))\delta_{\vec{\epsilon}'}(x,e'(t'))$
by $\delta_{\vec{\epsilon}}(e'(t'),e(t))
\delta_{\vec{\epsilon}'}(x,v)$ and send first $\vec{\epsilon}\to 0$.
Then 
$$
\delta_{\vec{\epsilon}}(e'(t'),e(t))\to
\frac{\delta(t)\delta(t')}{|\epsilon_{ab}
\dot{e}^a(0)\dot{e}^{\prime b}(0)|}
$$
and we can perform the integral. Since we are integrating over a square 
$[0,1]^2$ and
the two-dimensional delta-distribution is supported at a corner we pick up
a factor of $1/4$ upon setting $t=t'=0$ and dropping the integral. At 
last we send $\vec{\epsilon}'\to 0$.
\end{itemize}
\end{itemize}
Summarizing, we find ($V(\gamma)$ denotes the set of vertices of $\gamma$)
\be \label{24}
\hat{E}^i(x)f_\gamma
=-\frac{\hbar^2}{4\cdot 2}
\sum_{v\in V(\gamma)}\delta(x,v) \sum_{e,e'\in E(\gamma),e\cap e'=v}
\mbox{sgn}(e,e') \epsilon^{ijk}X^j_e X^k_{e'} f_\gamma
\ee
where $X^i_e:=X^i_e(0)$ is easily recognized as the right invariant 
vector field on $SU(2)$ evaluated at $g=h_e(0,1)$ and $\mbox{sgn}(e,e')$
is the sign of 
$\epsilon_{ab}\dot{e}^a(0)\dot{e}^{\prime b}(0)$ and so is an orientation 
factor. This furnishes the definition of the operator corresponding to 
the degeneracy vector.

We now will define the volume operator for any compact region 
$B\subset\Sigma$. Our first task is to define an operator corresponding
to $\det(q)$ and then to take its square root. Since $\det(q)$ is a 
density of weight two we expect this to be quite singular, in fact
the naive definition $\widehat{\det(q)}(x):=\hat{E}^i(x)\hat{E}^i(x)$ 
does not make any sense given the expression (\ref{24}) which involves 
a factor of $\delta(x,v)$. Thus we are lead to point-split the two 
degeneracy vector operators and to hope that 1) the regulated operator is
positive so that it makes sense to take its square root and 2) that
one can remove the regulator from the square root. Let us then define
similar as above
\be \label{25}
\widehat{\det(q)}_{\vec{\epsilon},\vec{\epsilon}'}(x)
:=\int_\Sigma d^2y\int d^2z 
\delta_{\vec{\epsilon}}(x,y)\delta_{\vec{\epsilon}'}(x,z)
\hat{E}^i(y)\hat{E}^i(z)
\ee
and apply it to a function cylindrical with respect to a graph $\gamma$.
Given (\ref{24}) the result is easily seen to be 
\ba \label{25a}
&&\widehat{\det(q)}_{\vec{\epsilon},\vec{\epsilon}'}(x) f_\gamma
\nonumber\\
&=&\frac{\hbar^4}{64}\sum_{v,v'\in V(\gamma)}
\delta_{\vec{\epsilon}}(x,v)\delta_{\vec{\epsilon}'}(x,v') \times\nonumber\\
&\times&\sum_{e_1,e_2\in E(\gamma),e_1\cap e_2=v} \;\;
\sum_{e_1',e_2'\in E(\gamma),e_1'\cap e_2'=v'} 
\mbox{sgn}(e_1,e_2) \mbox{sgn}(e_1',e_2') \times\nonumber\\
&\times&
[\epsilon^{ijk}X^j_{e_1} X^k_{e_2}] [\epsilon^{imn}X^m_{e_1'} X^n_{e_2'}]
f_\gamma\;.
\ea
We now will accomplish both hopes 1), 2) stated above by appropriately 
choosing the regulators.

1) Choose $\vec{\epsilon}=:\vec{\epsilon}'$, then
we are able to display (\ref{25}) as a 
square of an operator
\ba \label{26}
&&\widehat{\det(q)}_{\vec{\epsilon},\vec{\epsilon}}(x) f_\gamma
\nonumber\\
&=&\{\frac{\hbar^2}{8}\sum_{v\in V(\gamma)}
\delta_{\vec{\epsilon}}(x,v)\sum_{e_1,e_2\in E(\gamma),e_1\cap e_2=v} 
\mbox{sgn}(e_1,e_2) [\epsilon^{ijk}X^j_{e_1} X^k_{e_2}] \}^2
f_\gamma\;.
\ea
Since $X^i_{e_1} X^j_{e_2}$ commute for $e_1\not=e_2$ and because 
$i X_e^i$ is essentially self-adjoint with range in its domain, so is 
$X^i_{e_1} X^j_{e_2}$ and therefore the whole operator corresponding 
to one factor in (\ref{26}). Thus, (\ref{26}) is a square of 
essentially self-adjoint operators with range in its domain and so it 
is positive semi-definite. Therefore its square root is well defined.

2) Choose $\vec{\epsilon}$ small enough such that 
$\delta_{\vec{\epsilon}}(x,v)\delta_{\vec{\epsilon}}(x,v')=
\delta_{v,v'}[\delta_{\vec{\epsilon}}(x,v)]^2$, that is, given $\gamma,x$ 
we must choose $\vec{\epsilon}$ so small that for $v\not=v'$ not both
of them can be in the support of the function $\delta_{\vec{\epsilon}}(x,.)$
which is always possible. Then we may write (\ref{26}) as
\ba \label{27}
&&\widehat{\det(q)}_{\vec{\epsilon},\vec{\epsilon}}(x) f_\gamma
\nonumber\\
&=&\frac{\hbar^4}{64}\sum_{v\in V(\gamma)}
[\delta_{\vec{\epsilon}}(x,v)]^2
\{\sum_{e_1,e_2\in E(\gamma),e_1\cap e_2=v} 
\mbox{sgn}(e_1,e_2) [\epsilon^{ijk}X^j_{e_1} X^k_{e_2}] \}^2
f_\gamma, 
\ea
take its square root and define this to be the regulated operator 
corresponding to $\sqrt{\det(q)}$ : 
\be \label{28}
\widehat{\sqrt{\det(q)}}_{\vec{\epsilon}}(x) f_\gamma:=
\sqrt{\widehat{\det(q)}_{\vec{\epsilon},\vec{\epsilon}}(x)} f_\gamma\;.
\ee
In considering the limit $\vec{\epsilon}\to 0$ notice that for small enough
$\vec{\epsilon}$ at most one vertex of $\gamma$ lies in the support of
$\delta_{\vec{\epsilon}}(x,.)$. Therefore we can take the sum over 
vertices and the factor $[\delta_{\vec{\epsilon}}(x,v)]^2$ out of the 
square root and find that
\be \label{29}
\widehat{\sqrt{\det(q)}}_{\vec{\epsilon}}(x) f_\gamma=
\frac{\hbar^2}{8}\sum_{v\in V(\gamma)}
\delta_{\vec{\epsilon}}(x,v)
\sqrt{\{\sum_{e_1,e_2\in E(\gamma),e_1\cap e_2=v} 
\mbox{sgn}(e_1,e_2) [\epsilon^{ijk}X^j_{e_1} X^k_{e_2}]\}^2}
f_\gamma\;.
\ee
But now the limit $\vec{\epsilon}\to 0$ is trivial to take, we finally 
find that
\be \label{30}
\widehat{\sqrt{\det(q)}}(x) f_\gamma=
\frac{\hbar^2}{8}\sum_{v\in V(\gamma)}
\delta(x,v)\sqrt{\{\sum_{e_1,e_2\in E(\gamma),e_1\cap e_2=v} 
\mbox{sgn}(e_1,e_2) [\epsilon^{ijk}X^j_{e_1} X^k_{e_2}]\}^2}
f_\gamma
\ee
or in integrated form
\ba \label{31}
&&\hat{V}(B)f_\gamma:=[\int_B d^2x \widehat{\sqrt{\det(q)}}(x)]f_\gamma
\nonumber\\
&=& 
\frac{\hbar^2}{8}\sum_{v\in V(\gamma)\cap B}
\sqrt{\{\sum_{e_1,e_2\in E(\gamma),e_1\cap e_2=v} 
\mbox{sgn}(e_1,e_2) [\epsilon^{ijk}X^j_{e_1} X^k_{e_2}]\}^2}
f_\gamma\;.
\ea
Formula (\ref{31}) motivates to introduce the ``volume operator at a point"
$\hat{V}_v$ : For each integer $n\ge 2$ define $\{[v,n]\}$ to be the set of 
germs of $n$ analytical edges incident at $v$ (a germ of an analytical 
edge at a point $v$ is a complete set of analytical data available at 
$v$ that are necessary to reconstruct it, that is, essentially the 
coefficients of its Taylor series). For a germ 
$\vec{e}_n:=(e_1,..,e_n)\in\{[v,n]\}$ define 
$$\hat{V}_{\vec{e}_n}:=
\sqrt{\{\sum_{e_I,e_J\in\vec{e}} 
\mbox{sgn}(e_I,e_J) [\epsilon^{ijk}X^j_{e_I} X^k_{e_J}]\}^2}
$$
where the right invariant vector field $X_e^i(g)=X_e^i(gh)\forall h\in 
SU(2)$,
due to right invariance, depends really only on the germ of the edge $e$ 
because 
it acts on a function in the same way no matter how "short" the segment 
of $e$ is on which that function actually depends, as long as that 
segment starts at $v=e(0)$. In particular all $X_e^i$, $e$ incident at $v$, 
commute as long as their germs are different. Then 
\be \label{32}
\hat{V}(B)=\sum_{v\in B}\hat{V}_v \mbox{ where }
\hat{V}_v=\sum_{n=2}^\infty\sum_{\vec{e}_n\in\{[v,n]\}}\hat{V}_{\vec{e}_n}\;.
\ee
We see that $\hat{V}(B)$ is a densely defined, essentially self-adjoint,
positive semi-definite operator on $\cal H$ for each bounded region 
$B\subset\Sigma$. Its most interesting property is that it acts 
non-trivially only at vertices of the graph underlying a cylindrical 
function,
moreover, that vertex has to be such that at least two edges incident at 
it have linearly independent tangents there. This is in complete analogy 
with the volume operator of the 3+1 theory just that we need to replace 
everywhere valence three by valence two. Unlike the the three-dimensional
volume operator, however, its two-dimensional "brother" does not vanish at
two-valent and three-valent vertices at all as long as there are at least 
two edges with linearly independent tangents at the vertex under 
consideration. As we will see in the appendix, the two-dimensional 
volume operator is even positive definite on gauge invariant 
functions with two-and three valent vertices while the three-dimensional
volume operator annihilates such functions identically.
This is to be expected because by inspection of (\ref{32}) the principal 
symbol of that operator is non-singular on two-and three valent
vertices while in the three-dimensional case it is singular.\\
The fact that the volume operator acts only at vertices of the graph will 
enable us to take the infra-red limit in case we are dealing with 
asymptotically flat topologies and also ensure that the ultra-violet limit 
exists. Thus, the volume operator acts both as an IR and as an UV 
{\em dynamical} regulator, a point of view emphasized in \cite{TTMat}.\\
\\
Remark :\\
Notice that $q_{ab}=\{A_a^i,V\}\{A_b^i,V\}$ just as in the 
three-dimensional case. This observation lead in the three-dimensional case
to the construction of a length operator \cite{8}. The only crucial property
that was necessary to construct this operator was that the volume operator
acts only at vertices. Since that is true for the two-dimensional 
operator as well {\em we can therefore take over all the results and 
formulae from \cite{8} to the two-dimensional case}, except for the 
obvious differences which are due to different dimension and algebraic
expressions in terms of right invariant vector fields
of the volume operators. In particular, although the eigenvalues of the 
length operators are certainly different, qualitatively the spectrum is 
still discrete, the operator is positive semi-definite and essentially
self-adjoint and the length of a curve as measured by a spin-network 
state is different from zero only if at least one edge of the graph crosses
the curve, though not necessarily in a vertex. Thus we automatically have
a two-dimensional length operator as well.\\
The fact that the two-dimensional length operator is less degenerate than
the three-dimensional one can be traced back to the observation that
what is length in two dimensions is what is area in three dimensions.

\section{Regularization}

This section is divided into three parts : in the first part we will derive
a regulated Wheeler-DeWitt operator. The regularization consists in a 
triangulation of 
$\Sigma$ which is kept arbitrary at this stage. In the next part we will
specify the properties that we wish to impose on the triangulation and 
then make a particular choice which satisfies those properties.
Finally in the last part we complete the regularization by employing that 
triangulation and take the continuum limit which then equips us with a 
densely defined family of operators, one for each graph.\\
The presentation will be kept largely parallel to the one in \cite{2} in 
order to fasciliate comparison.

\subsection{Derivation of the regulated operator}

We wish to define an operator corresponding to
\ba \label{33}
H(N)&:=&\int_\Sigma d^2x N F_i\frac{E^i}{\sqrt{\det(q)}}\nonumber\\
&=&\frac{1}{2}\int d^2x N 
\epsilon^{ab}\epsilon_{ijk} F_i\frac{\{A_a^j,V\}\{A_b^k,V\}}{\sqrt{\det(q)}}
\nonumber\\
&=&\frac{1}{2}\int N \epsilon_{ijk} F_i\frac{\{A^j,V\}
\wedge\{A^k,V\}}{\sqrt{\det(q)}}\nonumber\\
&=&-\int N \mbox{tr}(F\frac{\{A,V\}\wedge\{A,V\}}{\sqrt{\det(q)}})
\ea
where we have used that $\mbox{tr}(\tau_i\tau_j\tau_k)=-\epsilon_{ijk}/2$
and (\ref{9}). Following the idea outlined in section 2 
consider now a point splitting of the above expression as follows : Let 
$\epsilon$ be a small number and $\chi_\epsilon(x,y):=
\theta(\frac{\epsilon}{2}-|x^1-y^1|)\theta(\frac{\epsilon}{2}-|x^2-y^2|)$
where $\theta(t)=1$ if $t>0$ and $0$ otherwise, that is, $\chi_\epsilon$ 
is the characteristic function of square of coordinate volume $\epsilon^2$ . 
Moreover, it is just 
true that $\{A_a^i(x),V\}=\{A_a^i(x),V(x,\epsilon)\}$ where 
$$
V(x,\epsilon):=\int_\Sigma d^2y \chi_\epsilon(x,y)\sqrt{\det(q)(y)}
$$
is the volume of the square around $x$ as measured by $q_{ab}$. Notice that
trivially $\lim_{\epsilon\to 0} V(x,\epsilon)/\epsilon^2=\sqrt{\det(q)(x)}$.
Therefore we have the identity (we write the density $F^i$ as a 
2-form) 
\ba \label{34}
&&\lim_{\epsilon\to 0}\int N(x) \mbox{tr}(F(x) \int \chi_\epsilon(x,y)
\frac{\{A(y),V\}\wedge\{A(y),V\}}{V(y,\epsilon)})\nonumber\\
&=&\lim_{\epsilon\to 0}\int N(x) \mbox{tr}(F(x) \int 
\frac{\chi_\epsilon(x,y)}{\epsilon^2}
\frac{\{A(y),V\}\wedge\{A(y),V\}}{V(y,\epsilon)/\epsilon^2})\nonumber\\
&=&\int N(x) \mbox{tr}(F(x) \int 
[\lim_{\epsilon\to 0}\frac{\chi_\epsilon(x,y)}{\epsilon^2}]
\frac{\{A(y),V\}\wedge\{A(y),V\}}
{\lim_{\epsilon\to 0}V(y,\epsilon)/\epsilon^2})\nonumber\\
&=&-\frac{1}{2}H(N),
\ea
that is, the point splitting singularity $1/\epsilon^2$ was {\em absorbed}
into $V(y,\epsilon)$.
The limit identity (\ref{34}) motivates to define a point split expression
\ba \label{35}
&&H_\epsilon(N):=-\int N(x) \mbox{tr}(F(x) \int \chi_\epsilon(x,y)
\frac{\{A(y),V\}\wedge\{A(y),V\}}{V(y,\epsilon)})\nonumber\\
&=&-\int N(x) \mbox{tr}(F(x) \int \chi_\epsilon(x,y)
\frac{\{A(y),V(y,\epsilon)\}}{\sqrt{V(y,\epsilon)}}\wedge
\frac{\{A(y),V(y,\epsilon)\}}{\sqrt{V(y,\epsilon)}})
\nonumber\\
&=& -4\int N(x) \mbox{tr}(F(x) \int \chi_\epsilon(x,y)
\{A(y),\sqrt{V(y,\epsilon)}\}\wedge\{A(y),\sqrt{V(y,\epsilon)}\}),
\ea
that is, the simple formula $\{.,\sqrt{V(y,\epsilon)}\}=\{.,V(y,\epsilon)\}/
(2V(y,\epsilon))$ enabled us to bring the volume functional from the 
denominator into the nominator, of course, inside the Poisson bracket.

The idea is now to replace Poisson brackets by commutators and the volume
functional by the volume operator and then take the limit $\epsilon\to 0$.
In order to do that we must first write (\ref{34}) in such quantities on 
which the volume operator knows how to act. Since, as obvious from the  
previous section, it only knows how to act on 
functions of holonomies along edges we must replace the connection 
field $A_a^i$ in (\ref{35}) by holonomies. We are thus forced to 
introduce a {\em triangulation} of $\Sigma$.\\
Denote by $\Delta$ a solid triangle. Single out one of the corners of the
triangle and call it $v(\Delta)$, the basepoint of $\Delta$. At $v(\Delta)$
there are incident two edges $s_1(\Delta),s_2(\Delta)$ 
of $\partial\Delta$ which we 
equip with outgoing orientation, that is, they start at $v(\Delta)$.
We fix the labelling as follows : let $s$ be the analytic extension 
of $s_1(\Delta)$ and $\bar{s}_1(\Delta)$ the half of $s$ starting at 
$v(\Delta)$ but not including $s_1(\Delta)-\{v\}$ with outgoing 
orientation at $v(\Delta)$. Let $U$ be a sufficiently small neighbourhood
of $v(\Delta)$ which is split into two halves by $s$. Define the {\em 
upper half} $U^+$ of $U$
to be that half of $U$ which one intersects as one turns $s_1(\Delta)$
counterclockwise into $\bar{s}_1(\Delta)$.
Now we require that there exists $U$ such that $U\cap s_2(\Delta)
=U^+\cap s_2(\Delta)$, that is, $s_2(\Delta)$ intersects the upper half of 
$U$. 
\begin{Definition} \label{def1}
Two analytical edges $e_1,e_2$ incident and outgoing at $v=e_1\cap e_2$ 
will be said to be right oriented iff
there exists a neighbourhood $U$ of $v$, its upper half $U^+$
being defined by $e_1$, such that $e_2$ intersects $U^+$.
\end{Definition}
This prescription is obviously 
diffeomorphism invariant. Notice that we did not, as it is usually done 
for triangulations, require that the tangents of the edges bounding $\Delta$
must have linearly independent tangents at their intersection. If they
are linearly independent then our prescription is equivalent to saying that
$\epsilon_{ab}\dot{s}_1^a(0)\dot{s}_2^b(0)>0$\\
Finally, let $a(\Delta)$ denote the remaining edge of $\partial(\Delta)$,
called the arc of $\Delta$, whose orientation we fix by requiring that it 
runs from the endpoint of $s_1(\Delta)$ to the endpoint of $s_2(\Delta)$.
Then $\partial\Delta=\alpha_{12}(\Delta)=s_1(\Delta)\circ a(\Delta)\circ
s_2(\Delta)^{-1}$ is called the loop of $\Delta$ based at $v(\Delta)$. We 
define also $\alpha_{21}(\Delta):=\alpha_{12}(\Delta)^{-1}$.

Let us now write the integral
over $\Sigma\times\Sigma$ in (\ref{35}) as a double sum of integrals
over $\Delta\times\Delta'$ where $\Delta,\Delta'$ are triangles of some
triangulation $T$ of $\Sigma$ 
\be \label{36}
H_{T,\epsilon}(N)=-4\sum_{\Delta,\Delta'\in T}
\mbox{tr}(\int_{\Delta'} N(x) F(x) \int_\Delta
\chi_\epsilon(x,y)
\{A(y),\sqrt{V(y,\epsilon)}\}\wedge\{A(y),\sqrt{V(y,\epsilon)}\})\;.
\ee
The purpose of the notation just introduced is that we may approximate, for 
sufficiently fine triangulation, each of the integrals by a function of 
holonomies as follows : Let $\delta$ be a small parameter and $s_i(\Delta)$
be the image of $[0,\delta]$ under the path $s_i(\Delta,t)$. Then, using 
smoothness of the connection we find
\ba \label{37}
&&N(v(\Delta'))\chi_\epsilon(v(\Delta'),y)\epsilon^{ij}
h_{\alpha_{ij}(\Delta')}\nonumber\\
&=& 2\delta^2 N(v(\Delta'))\chi_\epsilon(v(\Delta'),y)
\frac{\dot{s}_1^a(\Delta',0)\dot{s}_2^b(\Delta',0)}{2} F_{ab}(v(\Delta'))
+o(\delta^3)\nonumber\\
&=&2\int_{\Delta'} \chi_\epsilon(x,y)N(x)F(x)+o(\delta^3)\mbox{ and }
\nonumber\\ 
&&\chi_\epsilon(x,v(\Delta))\epsilon^{ij}
h_{s_i(\Delta)}\{h_{s_i(\Delta)}^{-1},\sqrt{V(v(\Delta),\epsilon)}\}
h_{s_j(\Delta)}\{h_{s_j(\Delta)}^{-1},\sqrt{V(v(\Delta),\epsilon)}\}
\nonumber\\
&=& \chi_\epsilon(x,v(\Delta))\delta^2
\dot{s}_1^a(\Delta,0)\dot{s}_2^b(\Delta,0)
\{A_a(v(\Delta)),\sqrt{V(v(\Delta),\epsilon)}\}
\{A_b(v(\Delta)),\sqrt{V(v(\Delta),\epsilon)}\})
\nonumber\\
&=& 2\int \chi_\epsilon(x,y)
\{A(y),\sqrt{V(y,\epsilon)}\}\wedge\{A(y),\sqrt{V(y,\epsilon)}\}
+o(\delta^3)
\ea
since the area of $\Delta$ is approximately $\delta^2\epsilon_{ab}
\dot{s}_1^a(\Delta,0)\dot{s}_2^b(\Delta,0)/2$
so that both integrals are of order $\delta^2$ provided that the 
tangents of $\partial(\Delta)$ at $v(\Delta)$ are linearly independent. 
Thus, up to an error of 
order $\delta^2$ which vanishes in the limit as the we remove the 
triangulation we may substitute (\ref{36}) by
\ba \label{38}
H_{T,\epsilon}(N)&=&-2\sum_{\Delta,\Delta'\in T}\epsilon^{ij}\epsilon^{kl}
N(v(\Delta'))\chi_\epsilon(v(\Delta'),v(\Delta))\times\nonumber\\
&\times& \mbox{tr}(h_{\alpha_{ij}(\Delta')}
h_{s_k(\Delta)}\{h_{s_k(\Delta)}^{-1},\sqrt{V(v(\Delta),\epsilon)}\}
h_{s_l(\Delta)}\{h_{s_l(\Delta)}^{-1},\sqrt{V(v(\Delta),\epsilon)}\})\;.
\nonumber\\ &&
\ea
The result (\ref{38}) is still purely classical and becomes $H(N)$ when
taking\\ 
1) first the continuum limit (that is, refining the triangulation ad 
infinitum) and \\
2) taking $\epsilon\to 0$ on smooth connections $A_a^i$ and 
smooth momenta $E^a_i$.\\ 
A second way to guide the limit and that leads to $H(N)$ 
is by ``synchronizing"
$\epsilon\approx\delta$ and to take $\delta\to 0$ as follows : for each 
$\Delta$ define $$
\epsilon(\Delta):=
\sqrt{|\epsilon_{ab}\dot{s}^a_1(\Delta,0)\dot{s}^a_2(\Delta,0)|}\delta,
$$
replace for each $\Delta'$ :\\
1)$\chi_\epsilon(v(\Delta),v(\Delta'))$ by 
$\chi_{\epsilon(\Delta')}(v(\Delta),v(\Delta'))$ 
and\\ 
2) $V(v(\Delta'),\epsilon)$ by $V(v(\Delta'),\epsilon(\Delta'))$\\
and then take $\delta\to 0$. Notice that this corresponds to introducing
$\epsilon(y)=\rho(y)\delta$ instead of $\epsilon$ in (\ref{34}) where 
$\rho(y)$ is an almost nowhere (with respect to $d^2x$)  
vanishing function such that $\rho(v(\Delta))\delta=\epsilon(\Delta)$.
Clearly $\rho$ must be almost nowhere vanishing as otherwise we do not
get a $\delta$ distribution in the limit $\delta\to 0$.
Notice that the set of $v(\Delta')$'s has $d^2x$ measure zero so that 
a vanishing $\rho(v(\Delta')$ is not worrysome.\\
It will be this latter limit which is meaningful in the quantum theory.\\
We have managed to write $H(N)$ in terms of holonomies 
up to an error which vanishes in either of the limits that we have indicated.

The next step is to turn (\ref{38}) into a quantum operator. This now just
consists in replacing $V(v(\Delta),\epsilon)$ by 
$\hat{V}(v(\Delta),\epsilon)$ and Poisson brackets by commutators times 
$1/(i\hbar)$ because we work in a connection representation. The result is
\ba \label{39}
\hat{H}_{T,\epsilon}(N)&=&\frac{2}{\hbar^2}\sum_{\Delta,\Delta'\in 
T}\epsilon^{ij}\epsilon^{kl}
N(v(\Delta'))\chi_\epsilon(v(\Delta'),v(\Delta))\times\nonumber\\
&\times& \mbox{tr}(h_{\alpha_{ij}(\Delta')}
h_{s_k(\Delta)}[h_{s_k(\Delta)}^{-1},\sqrt{\hat{V}(v(\Delta),\epsilon)}]
h_{s_l(\Delta)}[h_{s_l(\Delta)}^{-1},\sqrt{\hat{V}(v(\Delta),\epsilon)}])\;.
\ea
We wish to show that (\ref{39}) is densely defined in the limit 
$\epsilon\to 0$ no matter how we choose the triangulation $T$, as long as it 
is finite, thereby showing that 
the regulator $\epsilon$ can be removed {\em without encountering any 
singularity}. Thus, we prescribe the $\epsilon\to 0$ limit {\em before}
taking the limit of infinitely fine triangulation (continuum limit) and 
therefore have interchanged the order of limits as compared to the classical 
theory. However, as we will show shortly, one arrives at the same result
when synchronizing $\epsilon\approx\delta$ 
and taking $\delta$ sufficiently small but finite for the moment being which 
corresponds to 
the second way to guide the classical limit indicated above and therefore
interchanging the limits is allowed.\\
For that purpose let $f_\gamma$ 
be a function which is cylindrical with respect to a graph. Consider 
first some triangle $\Delta$ which does not intersect $\gamma$ at all.
Then it is easy to see that 
$$
[h_{s_l(\Delta)}^{-1},\sqrt{\hat{V}(v(\Delta),\epsilon)}]) f_\gamma=0\;
$$
The reason for this is that the graphs $\gamma$ and $\gamma\cup s_l(\Delta)$ 
then do not have any two-valent vertex in the box around 
$v(\Delta)$ parametrized by $\epsilon$ other than the vertices of 
$\gamma$ themselves. Thus the volume operator does not act on 
$h_{s_l(\Delta)}^{-1}$ and the commutator vanishes. It follows that 
only tetrahedra which intersect the graph contribute in (\ref{39}).
So let $\gamma\cap\Delta\not=\emptyset$. For the same reason as above we 
find a non-zero contribution only if $s_1(\Delta)$ or $s_2(\Delta)$ intersect
$\gamma$, that $a_{12}(\Delta)$ alone intersects $\gamma$ is not sufficient. 
Moreover, still 
for the same reason, if $s_i(\Delta)$ intersects $\gamma$ but not in the 
starting point of $s_i(\Delta)$ then we still get zero upon choosing 
$\epsilon$ sufficiently small so that the intersection point p lies outside 
the support of the characteristic function, that is,
$\chi_\epsilon(v(\Delta),p)=0$. Thus a triangle 
$\Delta$ contributes to 
(\ref{39}) if and only if $v(\Delta)\in\gamma$. But if that is true then
we may replace $\hat{V}(v(\Delta),\epsilon)$ by the operator 
$\hat{V}_{v(\Delta)}$ defined in (\ref{32}) and so the $\epsilon$-dependence
of $\hat{V}(v(\Delta),\epsilon)$ has dropped out. The remaining 
$\epsilon$-dependence now just rests in the function
$\chi_\epsilon(v(\Delta'),v(\Delta))$. Now, since we let $\epsilon\to 0$
first, at finite triangulation, we conclude altogether that the 
unrestricted double sum over triangles in (\ref{39}) collapses to a 
double sum over triangles subject to the condition that their basepoints 
coincide and lies on the graph. In formulae 
\ba \label{40}
&&\hat{H}_T(N)f_\gamma:=\lim_{\epsilon\to 0}\hat{H}_{T,\epsilon}(N)f_\gamma
\nonumber\\
&=&\frac{2}{\hbar^2}\sum_{\Delta,\Delta'\in 
T,v:=v(\Delta)=v(\Delta')\in\gamma} \epsilon^{ij}\epsilon^{kl}N(v)
\times\nonumber\\
&\times& \mbox{tr}(h_{\alpha_{ij}(\Delta')}
h_{s_k(\Delta)}[h_{s_k(\Delta)}^{-1},\sqrt{\hat{V}_v}]
h_{s_l(\Delta)}[h_{s_l(\Delta)}^{-1},\sqrt{\hat{V}_v}])f_\gamma
\ea
which displays $\hat{H}_T(N)$ as a densely defined operator which
does not suffer from any singularities because at finite triangulation 
there are only a finite number of terms involved in (\ref{40}), even
if $\Sigma$ is not compact.\\
Notice that in the $\epsilon\to 0$ limit we have recovered a gauge 
invariant operator as we should. \\
\\
Let us now show that one arrives at the same result by synchronizing
$\epsilon\approx\delta$ as above and taking $\delta$ sufficiently small 
but still finite : Namely, by choosing $\epsilon(\Delta')$ as above  
we have arranged that only the starting points of the $s_i(\Delta')$ are 
covered
by the $\epsilon(\Delta')$-box around $v(\Delta')$ that underlies the 
definition
of $\hat{V}(v(\Delta'),\epsilon(\delta'))$. This implies first of all 
that we need 
to sum only over $v(\Delta)=v(\Delta')$. Next, as we will be forced to 
adapt the triangulation to the graph anyway, we 
can arrange that the $\Delta$ intersect $\gamma$ only either in whole 
edges or in vertices of $\Delta$. If that is the case, then it follows that 
$[h_{s_i(\Delta)},\sqrt{\hat{V}(v(\Delta),\epsilon(\Delta))}]f_\gamma$ is 
non-vanishing only if $s_i(\Delta)$ intersects $\gamma$ in $v(\Delta)$
because the end-point is not covered by the $\epsilon(\Delta)$-box and 
if $s_i(\Delta)$ is contained in $\gamma$ but does not start in a 
vertex of $\gamma$ then the commutator vanishes due to the properties of the 
volume operator. This is enough to see that we arrive at (\ref{40}) again. \\
\\
In either way of taking the limit
we are now left with taking the continuum limit $\delta\to 0$ of refining 
the triangulation ad infinitum which we denote as $T\to\infty$. Certainly that
limit depends largely on the choice of the limit $T\to\infty$. For instance, 
if we are not careful and refine $T$ in such a way that the number of
basepoints of triangles that intersect $\gamma$ diverges we will not get
a densely defined operator. We see that we must choose $T$ according to
$\gamma$ so that we get actually a family of operators 
$$
\hat{H}_{\gamma,T}(N)=\hat{H}_{T(\gamma)}(N)
$$ 
where $T(\gamma)$ is a triangulation adapted to $\gamma$ together with a 
well-defined refinement procedure $T\to\infty$. We will propose such a 
$T(\gamma)$
in the next subsection guided by some physical principles. It will then
be our task to verify that the family $(\hat{H}_{\gamma,T}(N))$ still 
defines a linear operator.

\subsection{Choice of the triangulation}

So far everything what we said was in complete analogy with the 
three-dimensional case
\cite{2} except that there we did not even need a point-splitting. In 
particular, (\ref{40}) is the precise counterpart of the three-dimensional
Euclidean Hamiltonian constraint operator.

What is different now is that in the 3+1 case the volume operator was 
much more degenerate than in the 2+1 case, a result of which was that
a basepoint of a simplex had to coincide with a vertex of the graph in 
order to contribute without further specification of the triangulation. 
Therefore, it was sufficient to adapt the triangulation to the 
graph in such a way that, among other things, the number of simplices 
intersecting a vertex stays constant as one refines the triangulation in 
order to arrive at well-defined continuum limit. In the 2+1 case that is not
true any longer and one must worry about the number of 
triangles intersecting the graph $\gamma$ off the vertices of $\gamma$.

Let us adopt the physical principles listed in \cite{2} which should 
guide one of how to choose the triangulation. In brief, they were :
\begin{itemize}
\item[1)] 
The amount of ambiguity arising from the choice of the 
triangulation should be kept to a minimum.
\item[2)] 
The resulting operator should be non-trivial and not annihilate every state.
\item[3)]
The choice of the contributing $\Delta$ should be diffeomorphism 
covariant as to interact well with the diffeomorphism invariance of the 
theory.
\item[4)] 
The choice of the $\Delta$ should be canonical and not single
out one part of the graph as compared to the other or one graph as 
compared to another.
\item[5)]
The family of operators $\hat{H}_\gamma(N)$ should define a linear operator
$\hat{H}(N)$ (cylindrical consistency).
\item[6)] 
The resulting operator $\hat{H}_{\gamma,T}(N)$ should be densely 
defined with a well-defined continuum limit. That is, if $\Psi\in\Phi'$ is a 
diffeomorphism invariant distribution and $f_\gamma$ a function 
cylindrical with respect to a graph $\gamma$ then 
$$
\lim_{T\to\infty}\Psi(\hat{H}_{\gamma,T}(N)f_\gamma)=:
\Psi(\hat{H}(N)f_\gamma))
$$
exists. The fact that $\Psi$ is diffeomorphism invariant is because we 
actually want to define $\hat{H}(N)$ on solutions to the diffeomorphism 
constraint which turn out to be distributions \cite{5}
and so the above limit is
the precise sense in which $\hat{H}(N)$ is defined on distributions.
\item[7)]
The operator $\hat{H}(N)$ should be free of anomalies, that is, 
$$
\Psi([\hat{H}(M),\hat{H}(N)]\phi)=0
$$
for each $\phi\in\Phi$ and every diffeomorphism invariant $\Psi\in\Phi'$.
\end{itemize}
Since we wish to obtain a densely defined operator no matter how fine the
triangulation while keeping the extra structure coming from the 
triangulation to a minimum we are naturally lead to impose that the 
triangles that intersect $\gamma$ in its basepoint must be constant in 
number. There are only two diffeomorphism invariantly different 
possibilities : either $v(\Delta)$ is a vertex of $\gamma$ or it lies 
on an edge of $\gamma$ between its endpoints. Since we want to get a 
non-vanishing operator one of the two or both scenarios should happen.

Suppose first then that $v(\Delta)$ is an interiour point of an edge $e$.
Then there is no natural way how to choose the triangle $\Delta$ itself :
the only structure available is the edge $e$ and one may therefore choose 
one of $s_i(\Delta)$, say $s_1(\Delta)$, to lie 
entirely in $e$. But then $s_2(\Delta)$ should certainly not lie in $e$
otherwise $v(\Delta)$ would be a vertex of $\gamma\cup\partial\Delta$
with only co-linear tangents of edges incident at it and the volume operator
$\hat{V}_{v(\Delta)}$ would vanish. Thus there is at least a huge 
ambiguity in how to choose $s_2(\Delta)$.

If, on the other hand, $v(\Delta)$ is a vertex of the graph then there 
are at least two edges $e,e'$ of $\gamma$ incident at it and now it is a 
natural choice to assume that $s_i(\Delta)$ coincide with segments of
$e,e'$.\\
In conclusion, guided by the principle of introducing as few ambiguous 
elements as possible into the triangulation we are motivated to exclude
that a $v(\Delta)$ is an interiour point of an edge or that it anyway 
does not contribute. The latter can be achieved by assuming that the 
edges $s_i(\Delta)$ have co-linear tangents at $v$ in this case. 

Now we are left with those $v(\Delta)$ that are vertices of $\gamma$.
Following the principle that our prescription should be canonical we 
must have that either each vertex of $\gamma$ is a basepoint of some
$\Delta$ or none. Since the latter possibility is excluded by the 
principle of non-triviality we are now concerned with the issue of how many
$\Delta$'s should have basepoint in each $v\in V(\gamma)$. A natural 
answer to this question is that there should be as many such $\Delta$'s as
pairs of edges incident at $v$ because otherwise we would single out one 
pair to another. However, we still need to fulfill the requirement that 
the $\Delta$'s must come from a triangulation. Both observations motivate 
to define a whole family of triangulations adapted to $\gamma$ and to 
average over them.

Finally, we must fix in a diffeomorphism covariant way how to attach 
the arcs $a_{12}(\Delta)$ to $\gamma$. Notice that since $\Delta$ is a 
part of a triangulation with $v(\Delta)$ a vertex of $\gamma$ and with
$s_i(\Delta)$ segments of edges of $\gamma$ incident at $v$, it is 
possible that the endpoints of $a_{12}(\Delta)$ are actually basepoints 
of of other triangles $\Delta'$. This we either must avoid by choice
(which is possible) or we must impose that the tangents of $s_i(\Delta)$
and $a_{12}(\Delta)$ are co-linear at the endpoints of $a_{12}(\Delta)$.
As we will see, only the latter possibility leads to an anomaly-free theory.
This furnishes our preliminary investigation of how to choose $T(\gamma)$.

We will now prescribe $T(\gamma)$. The prescription is simpler but very
similar to the three-dimensional case.\\ 
Fix a vertex $v$ of $\gamma$ and let $n$ denote its valence. We can 
label the edges of $\gamma$ incident at $v$ in such a way that\\
1) the pairs $(e_1,e_2),(e_2,e_3),..,(e_{n-1},e_n)$ are right oriented 
and possibly also $(e_n,e_1)$ is right oriented according to definition
(\ref{def1}) and\\
2) as one encircles $v$ counter-clockwise, one does not cross any other 
edge after one crosses $e_i$ and before one crosses $e_{i+1}$ where
$e_{n+1}\equiv e_1$. We are 
going to construct a triangle $\Delta$ associated 
with each such right oriented pair which we will call $(e_1,e_2)$ from 
now on. We do not, in contrast to the 3+1 theory, construct a triangle 
associated with each pair because then $a_{12}$ in two dimensions would 
intersect 
not only $s_1,s_2$ but also other edges of the graph which we must avoid 
in order to have an anomaly-free theory as we will see. Moreover, in two 
dimensions the way we ordered the edges incident at $v$ is very natural 
and not available in three dimensions. \\
Finally, let $E(v)$ equal $n$ if $(e_n,e_1)$ is right oriented, 
otherwise let it equal $n-1$. In particular, for $n=2$ we must have 
$E(v)=1$.\\ 
We choose now 
$s_i(\Delta)$ to be any segment of $e_i$ which does not include the other 
endpoint of $e_i$ different from $v$ and which starts at $v$. Furthermore,
connect the endpoints of $s_1(\Delta)$ with the endpoint of 
$s_2(\Delta)$ by an arc $a_{12}(\Delta)$ with the special property 
that the tangent of $a_{12}(\Delta)$ is\\
1) parallel to the tangent of $s_1(\Delta)$
at the end-point of $s_1(\Delta)$ and \\
2) anti-parallel to the tangent of $s_2(\Delta)$
at the end-point of $s_2(\Delta)$.\\
Two remarks are in order :\\
a) Notice that we do not have to worry about any other edge of $\gamma$ 
intersecting $a_{12}(\Delta)$ because in two dimensions the topology 
of the routing of $a_{12}$ through the edges of $\gamma$ is very simple :
there is no way that $a_{12}$ can intersect any other edges of $\gamma$ 
other than $s_1,s_2$ given the labelling of $e_i$ made above. This is in 
contrast to the three-dimensional case where the topology of the routing
was extremely complicated to prescribe in a diffeomorphism covariant way.\\
b) In contrast to the three-dimensional case we here prescribed the 
$C^1$ properties of the edges $s_1,s_2,a_{12}$ at their intersection points.
The reason for this will become evident only later when we prove 
anomaly-freeness. We will see that the $C^1$ property of the 
intersection is crucial.

Whenever $(e_n,e_1)$ is a right oriented pair the $n$ triangles saturate
$v$. Otherwise there are only $n-1$ triangles and they do not yet 
saturate $v$. We follow the approach proposed in \cite{2} in order to 
achieve saturation. Namely, we take each of the $E(v)$ triangles and 
construct three more from it such that they altogether saturate $v$. Then 
we average over the $E(v)$ triangulations based on using only one such 
quadrupel of triangles. The details are as follows :\\
Let $s_i(t),a_{12}(t)$ be a parametrization of $s_i,a_{12}$ with $t\in 
[0,1]$. Let \\
$s_{\bar{i}}(t):=v-(s_i(t)-v)=2v-s_i(t)$,\\
$a_{\bar{1}\bar{2}}(t):=2v-a_{12}(t)$,\\
$a_{\bar{2} 1}(t):=s_{\bar{2}}(1)+t(s_1(1)-s_{\bar{2}}(1))$,\\
$a_{2\bar{1}}(t):=s_2(1)+t(s_{\bar{1}}(1)-s_2(1))$.\\
Then it is easy to see that $(s_{\bar{1}},s_{\bar{2}}),
(s_{\bar{2}},s_1),(s_2,s_{\bar{1}})$ are right oriented pairs and that 
the four triangles $\Delta_{12},\Delta_{\bar{1}\bar{2}},
\Delta_{\bar{2} 1},\Delta_{2\bar{1}}$ based on these triples of edges 
saturate $v$ (use $a_{12}(0)=s_1(1),a_{12}(1)=s_2(1)$ to see this).

Let now $S_i(v)$ denote the region in $\Sigma$ filled by these four triangles
based on a pair of edges $(e_i,e_{i+1})$ incident at $v$. Also denote by
$\Delta_i(v)$ the original triangle defined by $s_1,s_2,a_{12}$ for that 
pair from which we constructed the remaining three triangles as above.
Let $S(v):=\cup_{i=1}^{E(v)} S_i(v)$ be the union of these regions given 
by all the $E(v)$ pairs and let $\bar{S}_i(v)=S(v)-S_i(v)$. We will 
choose all the triangles so small that the $S(v)$ are mutually disjoint.
Finally, let $S=\cup_{v\in V(\gamma)}S(v)$ and $\bar{S}=\Sigma-S$. Then 
we can trivially decompose any integral over $\Sigma$ as follows
\ba \label{41}
\int_\Sigma&=&\int_{\bar{S}}+\sum_{v\in V(\gamma)}\int_{S(v)}
\nonumber\\
&=&\int_{\bar{S}}+\sum_{v\in V(\gamma)}\frac{1}{E(v)}\sum_{i=1}^{E(v)}
[\int_{\bar{S}_i(v)}+\int_{S_i(v)}]
\nonumber\\
&=&[\int_{\bar{S}}+\sum_{v\in V(\gamma)}\frac{1}{E(v)}\sum_{i=1}^{E(v)}
\int_{\bar{S}_i(v)}]
+\sum_{v\in V(\gamma)}\frac{1}{E(v)}\sum_{i=1}^{E(v)}\int_{S_i(v)}
\nonumber\\
&=&[\int_{\bar{S}}+\sum_{v\in V(\gamma)}\frac{1}{E(v)}\sum_{i=1}^{E(v)}
\int_{\bar{S}_i(v)}]
+[\sum_{v\in V(\gamma)}\frac{4}{E(v)}\sum_{i=1}^{E(v)}\int_{\Delta_i(v)}
+o(\delta^3)]\;.
\ea
In the last line we have exploited that for smooth integrands and small 
triangles the integral over each of the four triangles constructed is 
the same up to higher order in the parameter $\delta$ introduced before
equation (\ref{37}). It is clear that the term in the first square 
bracket of the last line in (\ref{41}) is a sum of integrals over regions of 
$\Sigma$ each of which does not contain vertices of $\gamma$. 

We are now ready to specify the family of triangulations $T(\gamma)$ of 
$\Sigma$ 
which by (\ref{37}) can actually be reduced to a family of triangulations of
$\bar{S},\bar{S}_i(v),\Delta_i(v)$ for $v\in V(\gamma), i=1,..,E(v)$ :\\ 
1) Triangulate $\Delta_i(v)$ by $\Delta_i(v)$\\
2) Triangulate $\bar{S}$ and $\bar{S}_i(v)$ arbitrarily subject to the 
condition that no basepoint of a triangle should lie on an edge of $\gamma$
or that all tangents at an intersection with an edge of $\gamma$ are 
co-linear\\ 
3) The triangles $\Delta_i(v)$ collapse to $v$ as $T\to\infty$ in such a 
way that all graphs $\gamma\cup\Delta_i(v)$ are diffeomorphic as 
$T\to\infty$. In fact as long as we keep the prescription of how to 
choose $s_i(\Delta),a_{12}(\Delta)$ specified above, all the graphs
$\gamma\cup\Delta$ are related by an analyticity preserving smooth 
diffeomorphism no matter how ``large" $\Delta$. Namely, such 
diffeomorphisms can leave the image of $\gamma$ invariant while putting
$a_{12}$ in any diffeomorphic shape. \\
Notice that now we have a well-defined prescription for the continuum limit
because by construction the triangles that triangulate $\bar{S},\bar{S}_i(v)$
do not contribute to the operator (\ref{40}). The fact that the number 
of triangles that have their basepoint in vertices of the graph (which 
are the only ones that contribute) stays constant (namely $E(v)$) 
indicates that the continuum operator will be densely defined.

\subsection{Continuum Limit}

Let us summarize : having specified the triangulation we have triangles 
$\Delta(\gamma,T)$ associated with the graph, more precisely $E(v)$ for each 
vertex $v$ of $\gamma$, the index $T$ indicating that the continuum limit 
has not been taken yet. Then the regulated operator (\ref{40}) becomes 
\ba \label{42} 
&&\hat{H}_T(N)f_\gamma:=\hat{H}_{\gamma,T}(N)f_\gamma:=
\frac{1}{\hbar^2}\sum_{v\in V(\gamma)}N(v)(\frac{4}{E(v)})^2
\sum_{v(\Delta),v(\Delta')=v}
\epsilon^{ij}\epsilon^{kl}\times\nonumber\\
&\times&\mbox{tr}(h_{\alpha_{ij}(\Delta')}
h_{s_k(\Delta)}[h_{s_k(\Delta)}^{-1},\sqrt{\hat{V}_v}]
h_{s_l(\Delta)}[h_{s_l(\Delta)}^{-1},\sqrt{\hat{V}_v}])f_\gamma
\ea
where we have dropped the dependence of the $\Delta$ on $\gamma,T$. Now, 
since as $T\to\infty$ all holonomies approach unity, the limit $T\to\infty$
does not have any meaning on the Hilbert space ${\cal H}=L_2(\agb,d\mu_0)$.
Indeed, on smooth connections we would get zero while on distributional 
connections the limit does not exist. Thus, the limit $T\to\infty$ must 
be understood in another way. Indeed, recall that we wanted to impose the 
Hamiltonian constraint actually on diffeomorphism invariant distributions 
$\Psi\in\Phi'$. Now, the operator $\hat{H}_T(N)$ defines for each 
$T$ an operator $(\hat{H}_T(N))'$ on $\Phi'$ by the equation 
\be \label{43}
[(\hat{H}_T(N))'\Psi](\phi):=\Psi(\hat{H}_T(N)\phi)\;\forall \Psi\in\Phi',\;
\phi\in\Phi
\ee
because $\hat{H}_T(N)$ has domain and range in $\Phi$ which is dense in 
$\cal H$, for each $T$. Now, if $\Psi$ is {\em diffeomorphism invariant} then
\be \label{44}
[(\hat{H}(N))'\Psi](f_\gamma):=\lim_{T\to\infty}[(\hat{H}_T(N))'\Psi](f_\gamma)
=\Psi(\hat{H}_{\gamma,T_0}(N)f_\gamma)
\ee
for each function $f_\gamma$ cylindrical with respect to a graph $\gamma$ 
and for each $\gamma$. In other words, the number 
$\Psi(\hat{H}_{\gamma,T}(N)f_\gamma)$ does not change under variation of 
$T$ which by prescription corresponded to a diffeomorphism and so on 
diffeomorphism invariant states we may evaluate it on any finite value $T_0$
and the $T\to\infty$ limit is trivial. It follows that on diffeomorphism 
invariant states the continuum limit is already taken for 
$(\hat{H}_T(N))'$.\\
In fact, it is easy to see that this result can be extended to any product
\\
$[\hat{H}_T(N_1)\hat{H}_T(N_2)..\hat{H}_T(N_n)]'$ because the triangles 
attached have, at each level, an unambiguously defined diffeomorphism 
covariant location. This observation is needed in order to give sense to
commutator computations \cite{2,9}.\\  
In the sequel we will drop the index $T$ and understand that when finally
evaluating everything on diffeomorphism invariant distributions the 
value of $T$ is irrelevant.

\section{Consistency}

There are two kinds of consistencies to be discussed :\\
The first is the cylindrical consistency, that is, we have obtained 
a family of operators $(\hat{H}_\gamma(N))_\gamma$ which should be 
projections to cylindrical subspaces of a ``mother" $\hat{H}(N)$.
That such a $\hat{H}(N)$ exists has to be proved.\\
The second is that we need to make sure that $\hat{H}(N)$ does not 
suffer from quantum anomalies.

\subsection{Cylindrical Consistency}

In proving that a family of operators $(\hat{O}_i,D_i)_{i\in I}$ on a 
Hilbert space $\cal H$, where $D_i$ is the domain of $\hat{O}_i$
and where $I$ is some 
partially ordered index set $I$ with ordering relation $<$, is cylindrically 
consistent we need to reveal that whenever $i<j$ that $\hat{O}_j$ is an 
extension of $\hat{O}_i$, that is\\
1) The domain of $\hat{O}_i$ is contained in that of $\hat{O}_j$, 
$D_i\subset D_j$ and \\
2) The restriction of $\hat{O}_j$ to $D_i$ coincides with $\hat{O}_i$,
$(\hat{O}_j)_{|i}=\hat{O}_i$.\\
Let us check that this is the case for our operator family. Recall that
a spin-network state depends on all of its edges non-trivially in the 
sense that all edges carry spin $j>0$. The space $\Phi_\gamma$ is the set
of finite linear combinations of spin-network states which depend on the 
graph $\gamma$. Now, while the set of graphs can be partially ordered by 
the inclusion relation, the set of cylindrical functions cannot because 
a function which is defined on a smaller graph is defined also on any 
bigger graph that properly contains it, however, the additional edges in 
that graph automatically carry spin zero and so the cylindrical subspaces 
cannot be compared. Another way of saying this is that given a 
cylindrical function $f$ we can uniquely decompose it as 
$f=\sum_\gamma f_\gamma,\; f_\gamma\in\Phi_\gamma$ and on $f_\gamma$ we 
have unambiguously $\hat{H}(N)f_\gamma=\hat{H}_\gamma(N)f_\gamma$. We 
cannot write $\hat{H}(N)f_\gamma=\hat{H}_{\gamma'}(N)f_\gamma$ with
$\gamma\subset\gamma'$ because there is a condition on the spins of the 
edges of $\gamma'$ involved when applying $\hat{H}_{\gamma'}$ which is 
not satisfied for $f_\gamma$. In other words, $\Phi_\gamma\cap\Phi_{\gamma'}
=\emptyset$ if $\gamma\not=\gamma'$\\
We conclude that the family $(\hat{H}_\gamma(N))$ is trivially
cylindrically consistently defined and therefore defines a linear operator
on all of $\cal H$.

\subsection{Anomaly-freeness}

Recall that the classical Dirac algebra is given by
$$
\{H(M),H(N)\}=\int_\Sigma d^2x (M{,a} N-M N_{,a})q^{ab} V_b
$$
where $V_a$ is the vector constraint. That is, the Poisson bracket 
between two Hamiltonian constraints evaluated on the constraint surface 
defined by the diffeomorphism constraint vanishes.

In the quantum theory one would therefore like to verify that naively\\
$[\hat{H}(M),\hat{H}(N)]f=0$ for any state $f$ that satisfies $\hat{V}_a f=0$.
Several subtleties arise :
\begin{itemize}
\item[1)] 
The solutions $\hat{V}_a f$ are in general no elements of the Hilbert 
space but generalized eigenvectors (distributions). Indeed, in this 
context the solutions of the diffeomorphism constraint are not elements of
$\cal H$ but of $\Phi'$ where we have the proper inclusion
$\Phi\subset{\cal H}\subset\Phi'$. Thus, since $\hat{H}(N)$ is defined only
on $\Phi$, the only operator that is defined on $\Phi'$ is the dual
$(\hat{H}(N))'$ via the pairing 
$\Psi(\hat{H}(N)\phi)=[(\hat{H}(N))'\Psi](\phi)$.
\item[2)] 
Observe that the operator $(\hat{H}(N))'$ was not defined on every
distribution but actually only on those that are solutions of the 
diffeomorphism constraint. Now even if $\Psi$ is diffeomorphism invariant,
that is, 
$\Psi(\hat{U}(\varphi)f_\gamma):=\Psi(f_{\varphi(\gamma)})=\Psi(f_\gamma)$,
then $(\hat{H}(N))'\Psi$ is not any longer as one can easily check. Thus 
we cannot verify that $[(\hat{H}(M))',(\hat{H}(N))']\Psi=0$, this equation 
is simply not defined. However, what {\em is} well-defined is 
$([\hat{H}(M),\hat{H}(N)])'\Psi=0$ and this is what we are going to verify.
Indeed, there is no hope to make sense out of 
$[(\hat{H}(M))',(\hat{H}(N))']\Psi$ since not even classically $H(M)$ is
diffeomorphism covariant. On the other hand, one could proceed as in \cite{9}
and {\em define} $\hat{H}'(M)\hat{H}'(N):=(\hat{H}(N)\hat{H}(N))'$ which
makes sense again on diffeomorphism invariant states.
\item[3)] 
One might be even more ambitious and ask that 
\be \label{44a}
([\hat{H}(M),\hat{H}(N)])'=
(\widehat{\int_\Sigma d^2x (M_{,a} N-M N_{,a})q^{ab} V_b})'
\ee
that is, the Dirac algebra is faithfully implemented in the quantum theory.
However, there are several issues that prevent us from doing so. First of 
all, the generator of diffeomorphisms, $V_a$, does not have a quantum 
analogue, the diffeomorphism group does not act strongly continuously on
$\cal H$. So the only thing that we can hope to obtain is something 
like $\hat{O}'[\hat{U}(\varphi)-1]$ for the right hand side of 
(\ref{44}) for some $\varphi\in\mbox{Diff}(\Sigma)$ and some dual operator
$\hat{O}'$ (notice that $\hat{U}(\varphi)'=\hat{U}(\varphi^{-1})$ can be 
extended 
to all of $\Phi'$). Secondly, the situation is even worse for $q^{ab}$. 
Thirdly,
since, as we said, $([\hat{H}(M),\hat{H}(N)])'$ is only well-defined on
diffeomorphism invariant distributions, then either the dual of the 
commutator vanishes or it does not. In the latter case there is an anomaly
even in the sense of $\hat{U}(\phi)-1$. In the former case we get just 
zero but then we can trivially make an equality of the form 
$$
([\hat{H}(M),\hat{H}(N)])'=\hat{O}'[\hat{U}(\varphi)-1]
$$
for any $\hat{O}'$ that we like. It then remains to ask whether one can 
somehow make sense out of an operator corresponding to the {\em combination}
$\int_\Sigma d^2x (M_{,a} N-M N_{,a})q^{ab} V_b$ and that
is actually the case : We will not prove this assertion here but refer 
the reader to \cite{9} which treats the 3+1 case but from which
it is obvious that the result can be extended to the 2+1 case.
\end{itemize}
Summarizing, we will check that $([\hat{H}(M),\hat{H}(N)])'\Psi=0$
on diffeomorphism invariant states. The key element of the proof is the 
following : as is obvious from (\ref{42}), if $f_\gamma$ is a function 
cylindrical with respect to a graph, then $\hat{H}(N)f_\gamma$ is a 
linear combination of functions each of which depends on graphs with new 
vertices not
contained in $\gamma$. More precisely, if $v\in V(\gamma)$ then 
for each triangle $\Delta$ based at $v$ there is a term 
$\frac{4N_v}{E(v)}\hat{H}_\Delta f_\gamma$ and this function is a linear 
combination of functions $f'$ each of which depends on a graph 
$\gamma'$ contained in the following list :\\
$\gamma\cup\Delta,(\gamma\cup\Delta)-s_1(\Delta),
(\gamma\cup\Delta)-s_2(\Delta),
(\gamma\cup\Delta)-(s_1(\Delta)\cup s_2(\Delta))$. Whether they 
appear depends on the spins 
of the graph $\gamma$. In any case these functions $f'$ depend on two more 
vertices $v_1,v_2$ coming from the endpoints of the arc $a_{12}(\Delta)$.
They may not depend on the original vertex $v$ if that vertex was 
two-valent with spins of the edges $e_i$ corresponding to $s_i(\Delta)$ being
$j=1/2$ for both $i=1,2$. In that case 
$[h_{s_i(\Delta(v))}^{-1},\hat{V}_v]f'=0$ because 
neither $f'$ nor $h_{s_i(\Delta(v))}^{-1}f'$ depend on graphs with more 
than one edge incident at $v$.\\ 
The point is now that $[h_{s_i(\Delta(v_1))}^{-1},\hat{V}_{v_1}]f'=
[h_{s_i(\Delta(v_2))}^{-1},\hat{V}_{v_2}]f'=0$. The reason for this is that
the vertices $v_1,v_2$ in the graphs on which $f'$ and 
$h_{s_i(\Delta(v))}^{-1}f'$ depend does not have edges with linearly 
independent tangents incident at it so that the volume operator annihilates
these functions.\\
Let us now write (\ref{42}) in the form 
\ba \label{45} 
\hat{H}_\gamma(N)&=&
\frac{32}{\hbar^2}\sum_{v\in V(\gamma)}N(v)\hat{H}_{\gamma,v}
\nonumber\\ 
\hat{H}_{\gamma,v}&=&\frac{1}{E(v)^2}
\sum_{v(\Delta),v(\Delta')=v} \hat{H}_{\gamma,v,\Delta,\Delta'}\nonumber\\
\hat{H}_{\gamma,v,\Delta,\Delta'}&=&
\epsilon^{ij}\epsilon^{kl}
\mbox{tr}(h_{\alpha_{ij}(\Delta')}
h_{s_k(\Delta)}[h_{s_k(\Delta)}^{-1},\sqrt{\hat{V}_v}]
h_{s_l(\Delta)}[h_{s_l(\Delta)}^{-1},\sqrt{\hat{V}_v}])
\ea
The function $\hat{H}_{\gamma,v}f_\gamma$ now can be written as a linear 
combination of functions $f'_{\gamma'}$ each of which depends on a graph 
$\gamma'$ which is a proper subgraph of the graph 
$\gamma(v):=\gamma\cup_{v(\Delta)=v(\Delta')=v}[\Delta\cup\Delta']$ and 
we will mean by $\hat{H}_{\gamma(v),v'}$ the operator that reduces to 
$\hat{H}_{\gamma',v'}$ on $f_{\gamma'}$ for each $v'\in V(\gamma')$ and 
is zero if $v'\not\in V(\gamma')$. \\
With this preparation we evaluate
\ba \label{46}
&&[\hat{H}(M),\hat{H}(N)]f_\gamma=
\sum_{v\in V(\gamma)}[N_v \hat{H}(M)-M_v \hat{H}(N)]
\hat{H}_{\gamma,v}f_\gamma\nonumber\\
&=& \sum_{v\in V(\gamma)}\sum_{v'\in V(\gamma(v))}
[N_v M_{v'}-M_v N_{v'}] 
\hat{H}_{\gamma(v),v'}\hat{H}_{\gamma,v}f_\gamma
\nonumber\\
&=& \sum_{v,v'\in V(\gamma)}
[N_v M_{v'}-M_v N_{v'}] 
\hat{H}_{\gamma(v),v'}\hat{H}_{\gamma,v}f_\gamma
\nonumber\\
&=& \frac{1}{2}\sum_{v,v'\in V(\gamma)}
[N_v M_{v'}-M_v N_{v'}] 
[\hat{H}_{\gamma(v),v'}\hat{H}_{\gamma,v}
-\hat{H}_{\gamma(v'),v}\hat{H}_{\gamma,v'}]f_\gamma\;.
\ea
Here we have used our notation to write the commutator as a double sum
in $v\in V(\gamma),v'\in V(\gamma(v))$ in the second line, then in the 
third line we have used the important fact that the constraint does not
act at the new vertices that it creates so that the sum over $v'\in 
V(\gamma(v))$ collapses to a sum over the original $v'\in V(\gamma)$ and 
in the last step we have used the antisymmetry in the lapse functions to
write the product of operators as their antisymmetrized sum of products.
Clearly the term with $v=v'$ vanishes trivially. If $v\not=v'$ then, 
since $\hat{H}_{\gamma,v}$ manipulates the graph only in a small 
neighbourhood of $v$, we can commute the two operators in the last line
of (\ref{46}) to write both with the vertex $v$ to the right hand side
as
\be \label{47}
[\hat{H}(M),\hat{H}(N)]f_\gamma=
\frac{1}{2}\sum_{v,v'\in V(\gamma)}
[N_v M_{v'}-M_v N_{v'}] 
[\hat{H}_{\gamma(v),v'}\hat{H}_{\gamma,v}
-\hat{H}_{\gamma,v'}\hat{H}_{\gamma(v'),v}]f_\gamma\;.
\ee
Now by inspection of (\ref{45}) we see that the last square bracket 
is a linear combination of functions of the type $f-f'$ where $f,f'$ are 
related by an analyticity preserving diffeomorphism by construction of 
the triangulation which relates different choices for the loop attachment by
such a diffeomorphism (this point is explained in more detail in 
\cite{2}). Thus when evaluating (\ref{47}) on a diffeomorphism invariant
state we can remove those diffeomorphisms and obtain just zero.\\
This suffices to show $([\hat{H}(M),\hat{H}(N)])'=0$.

Notice that it was essential in the argument that the additional vertices 
created by $\hat{H}(N)$ when acting on $f_\gamma$ do not contribute as we 
showed. If that was not the case the commutator would not vanish on 
diffeomorphism invariant distributions which is why we attached the loop 
in such a particular, $C^1$, way.

\section{Solving the theory}

This section is divided into two parts : In the first part we will 
describe the complete space of solutions to both the diffeomorphism and 
Hamiltonian constraint. In the second part this solution space is 
shown to contain the solutions to the curvature constraint which can be 
formulated in our language as well \cite{10}. One can equip the solution 
space with at least two very natural inner products. 
One of them is the inner product appropriate for the curvature 
constraint, the other one arises from direct construction of the 
solutions in the first part this section. Neither of these inner 
products give all solutions a finite norm.

\subsection{Complete set of solutions to all constraints}

Let be given a spin-network state $T_{\gamma,\vec{j},\vec{c}}$ and let
$$\{T_{\gamma,\vec{j},\vec{c}}\}:=
\{\hat{U}(\varphi)T_{\gamma,\vec{j},\vec{c}},\;
\varphi\in\mbox{Diff}(\Sigma)\}$$
be its orbit under the diffeomorphism group of analyticity preserving smooth 
diffeomorphisms. We define a diffeomorphism invariant distribution on $\Phi$ 
by $$
[T_{\gamma,\vec{j},\vec{c}}]:=\sum_{T\in\{T_{\gamma,\vec{j},\vec{c}}\}} T\;.
$$
That this is a continuous linear functional on $\Phi$ follows from 
the fact that the spin-network states form an orthonormal basis
by the argument given in \cite{9}. Therefore
It is also clear that every diffeomorphism 
invariant state is a linear combination of such
$[T_{\gamma,\vec{j},\vec{c}}]$'s so that by this procedure we can claim
to have found the general solution $\Psi$ to the diffeomorphism constraint
$\Psi(\hat{U}(\varphi)f)=\Psi(f)\;\forall\varphi\in\mbox{Diff}(\Sigma)$. 
We will call this space $\Phi'_{Diff}$\\
Given any $f\in\Phi$ we can uniquely decompose it as 
$f=\sum_I f_I T_I$ where $f_I$ are some constants and $T_I$ are 
spin-network states. We then define $[f]:=\eta_{Diff}f:= \sum_I f_I [T_I]$. 
Notice that
one cannot define $[f]$ as the sum of all states which are in its orbit
under diffeomorphisms since spin-network states defined on different 
graphs have uncountably infinite sets of diffeomorphisms that move one 
graph but not another. We have been here imprecise with the issue
of graph symmetries which alter the above formulae somewhat. See \cite{9}
for more details.\\
This construction can be used to define an inner product on $\Phi'_{Diff}$
by 
$$
<[f],[g]>_{Diff}:=[f](g)
$$ 
which is clearly a positive definite 
sesquilinear form and equips $\Phi'_{Diff}$ with the structure of a 
pre-Hilbert space. \\
\\
Remark :\\
It has been shown in \cite{5} that {\em if} there are only strongly 
diffeomorphism invariant observables in the theory {\em then} 
those observables define a superselection rule, namely, they cannot map
between spin-network states based on graphs which are in different 
diffeomorphism equivalence classes. As a result, the group average could 
be defined differently in every sector, that is, the inner product in 
every sector can be chosen individually which amounts to the ambiguity 
that the particular way of averaging given by $[f]$ is 
not selected by physical principles, meaning that for every 
diffeomorphism equivalence class of graphs there could be a different 
constant that multiplies $[T_{\gamma,\vec{j},\vec{c}}]$ in $[f]$. \\
However, as there are clearly weakly diffeomorphism invariant observables 
which, together with the Hamiltonian constraint map between those sectors,
there is no superselection rule and the way we have averaged is selected 
by the requirement that averaged spin-network states remain orthonormal
\cite{9}. \\
\\
We now wish to employ this result to find the general solution to all 
constraints. To that end, consider the set
$$
R:=\{\hat{H}(N)\phi,\; N\in {\cal S},\;\phi\in\Phi\},
$$
the range of the Hamiltonian constraint on $\Phi$ where ${\cal S}$ 
denotes the usual Schwartz space of test functions of rapid decrease.
Consider its orthogonal complement in $\Phi$ denoted $S:=R^\perp\subset\Phi$. 
Finally, consider the set $\{[s],\; s\in S\}$. Then it is 
easy to see that every solution to all constraints is a linear combination
of elements of this set and we will call the resulting span $\Phi'_{phys}$. 
Namely, let $s=\sum_I s_I T_I\in S$ then 
by definition $\sum_I \bar{s}_I f_I=0$ for any $f=\sum_I f_I T_I\in R$.
Thus $[s](f)=\sum_I \bar{s}_I [T_I](f)=\sum_I \bar{s}_I f_I=0 \forall 
f\in R$.

A geometrical construction of the space $S$ was given for the 
three-dimensional theory in \cite{2}. Here we could proceed similarly.
However, since this is only a model we restrict ourselves to showing
that the space $\Phi'_{phys}$ is uncountably infinite dimensional. Namely,
a particular simple class of vectors in $S$ consists of those elements of 
$\Phi$ which are linear combinations of spin-network states whose 
underlying graph is not of the form
$\gamma\cup\a_{12}(\Delta),
[\gamma\cup\a_{12}(\Delta)]-s_1(\Delta),
[\gamma\cup\a_{12}(\Delta)]-s_2(\Delta),
[\gamma\cup\a_{12}(\Delta)]-[s_1(\Delta)\cup s_2(\Delta)]$ for any 
$\Delta=\Delta(\gamma)$, $v(\Delta)\in V(\gamma)$ and $\gamma$ is a
graph underlying the same restriction but is otherwise {\em arbitrary}.
This particular class of solutions has the property that all of the 
resulting $[s]$ are normalizable with respect to $<.,.>_{Diff}$ 
while genuine elements of $\Phi'_{phys}$ will
not be normalizable with respect to the kinematical inner product on 
$\Phi'_{Diff}$. On the other hand, since the 2+1 Hamiltonian constraint
really resembles the 3+1 Euclidean Hamiltonian constraint it follows from 
the redults on the kernel of the Euclidean Hamiltonian constraint given in 
\cite{2,9} that {\em every} solution is a (possibly infinite) linear 
combination of basic solutions each of which is in fact {\em normalizable}
with respect to $<.,.>_{Diff}$. \\
We will see that the solutions to the curvature 
constraint are not normalizable with respect to $<.,.>_{Diff}$ and one needs 
to define another appropriate inner product $<.,.>_{curv}$ on the subset
of $\Phi'_{phys}$ corresponding to the solutions of the curvature 
constraint. However, it will turn out that the natural inner product 
$<.,.>_{phys}$ for our Hamiltonian constraint as suggested by \cite{9} is 
such that curvature constraint solutions are still not normalizable and
so $<.,.>_{curv}$ and $<.,.>_{phys}$ define genuinely non-isometric 
Hilbert spaces. We will turn to that issue in the next subsection.

\subsection{Comparison with the Topological Quantum Field Theory}

As shown in \cite{10}, in our language a solution to the curvature 
constraint
$F_i=0$ in the quantum theory is a distribution $\Psi_f\in\Phi'$ given by
$\Psi_f:=\delta_{\mu_0}(F)f$ for any $f\in \Phi$. Here $\delta_{\mu_0}(F)$
is a $\delta$ distribution with respect to the inner product on ${\cal H}$
which has support on the space of flat connections modulo gauge 
transformation $\cal M$. More precisely, we have the following : 
Any function on $\cal M$ is a gauge invariant function which depends 
on the connection only through the holonomies along (representants of) the 
independent generators $\alpha_1,..,\alpha_n$ of the fundamental group
$\pi_1(\Sigma)$, that is, 
$f(A_0)=f_n(h_{\alpha_1}(A_0),..,h_{\alpha_n}(A_0))$
for $A_0\in{\cal M}$. The 
measure $\nu_0$ on $\cal M$ for gauge group $G$ is defined by 
$$
\int_{{\cal M}} d\nu_0(A_0) f(A_0):=\int_{G^n} d\mu_H(g_1)..d\mu_0(g_n)
f_n(g_1,..,g_n)
$$
where $\mu_H$ denotes the Haar measure on $G$. Then the delta distribution
for flat connections is given by
\be \label{48}
\delta_{\mu_0}(F(A)):=\int_{{\cal M}}d\nu_0(A_0) \delta_{\mu_0}(A_0,A)
\ee
where
\be \label{49}
\delta_{\mu_0}(A_0,A)=\sum_{\gamma,\vec{j},\vec{c}}
\overline{T_{\gamma,\vec{j},\vec{c}}(A)}
T_{\gamma,\vec{j},\vec{c}}(A_0)
\ee
and the sum runs over {\em all} possible spin-network states. It is 
possible to arrive at (\ref{48}) from first principles by following 
the group average proposal \cite{10}.\\
It is also possible to write (\ref{48}) as a linear combination of 
distributions in $\Phi'_{Diff}$. To that 
end, denote by $I$ the label of a spin-network state and define
$T_{[I]}:=[T_I]$. Notice that the integral 
$k_I:=\int_{{\cal M}} d\nu_0(A_0)T_I(A_0)=:k_{[I]}$ is 
diffeomorphism invariant and thus only depends only on $[I]$. Then we may 
write 
\be \label{50}
\delta_{\mu_0}(F(A))=\sum_{[I]} k_{[I]} \overline{T_{[I]}(A)}\;.
\ee
It is easy to see \cite{10} that (\ref{50}) is a distribution on 
$\Phi'_{Diff}$ and certainly it is a distribution on $\Phi$.
However, (\ref{50}) is not normalizable with respect to 
$<.,.>_{Diff}$ :\\ 
To see this we use (\ref{50}) to notice that we can write
$\delta(F(A))=\eta_{Diff} f_F$ where $f_F:=\sum_{[I]} \overline{k_{[I]}}
T_{I_0([I])}(A)$ and $I_0([I])\in [I]$ is an arbitrary choice.
Thus by definition of the inner product between diffeomorphism invariant 
distributions we find 
\be \label{diverge} 
||\delta_{\mu_0}(F)||_{Diff}^2=(\eta_{Diff}f_F)(f_F)=\sum_{[I]} 
|k_{[I]}|^2
\ee
where the sum is over diffeomorphism equivalence classes of spin-network 
labels. But quantity (\ref{diverge}) is just plainly infinite :\\
Namely, it follows from the definition of $k_I=k_{[I]}$
that $k_{[I]}=T_I(A=0)$ whenever the graph underlying $I$ is contractable.
There are an at least countably infinite number of contractable,
mutually non-diffeomorphic, non-trivial graphs $\gamma_n,\; n=1,2,..$
in any $\Sigma$. An example is given by choosing $\gamma_n$ to be an
$n$-link, that is, a union of $n$ mutually non-intersecting loops 
$\alpha_1,..,\alpha_n$
homeomorphic to a circle each of which is homotopically trivial 
(contractable). Choose $I_n$ such that 
$T_{I_n}(A)=\prod_{k=1}^n T_{\alpha_k}(A)$ where 
$T_\alpha(A):=\mbox{tr}(h_\alpha(A))$ is the Wilson-Loop function and 
$h_\alpha(A)$ denotes the holonomy of $A$ along the loop $\alpha$. 
Using the basic integral $\int_{SU(2)} d\mu_H(g) \bar{g}_{AB} g_{CD}
=\frac{1}{2} \delta_{AC}\delta_{BD}$ it is 
easy to see that $T_{I_n}$ provide an orthonormal system of spin-network
states. But $T_{I_n}(A=0)=2^n$ and so (\ref{diverge}) contains the 
meaningless sum $\sum_{n=1}^\infty 2^n$.\\
\\
We must check whether or not $\Psi_f$ is also a solution to the constraint
$\hat{H}(N)$ (it obviously is diffeomorphism in variant). To that end we must
compute
\ba \label{51}
\Psi_f(\hat{H}(N) f_\gamma)&=&\int_\agb d\mu_0 \delta_{\mu_0}(F(A))
(\overline{f}\hat{H}_\gamma(N)f_\gamma)(A)\nonumber\\
&=&\int_{{\cal M}} d\nu_0(A_0)(\overline{f}\hat{H}_\gamma(N) f_\gamma)(A_0)
=0
\ea 
because either $\hat{H}_\gamma(N) f_\gamma$ is identically zero or it is 
a linear combination of the vectors (recall (\ref{45}))
$$
\hat{H}_{\gamma,v,\Delta,\Delta'}f_\gamma
=-2\epsilon^{ij}\epsilon^{kl}
\mbox{tr}(h_{\alpha_{ij}(\Delta')}\tau_m)
\mbox{tr}(\tau_m h_{s_k(\Delta)}[h_{s_k(\Delta)}^{-1},\sqrt{\hat{V}_v}]
h_{s_l(\Delta)}[h_{s_l(\Delta)}^{-1},\sqrt{\hat{V}_v}])f_\gamma
$$
which therefore are proportional to the matrix elements of
$[h_{\alpha_{12}}-h_{\alpha_{12}}^{-1}]$ for a contractable loop $\alpha$ 
which 
vanishes on $A_0\in{\cal M}$. Here we have used the $su(2)$ Fierz identity 
$\mbox{tr}(\tau_i A)\mbox{tr}(\tau_i B)=\mbox{tr}(A)\mbox{tr}(B)/4
-\mbox{tr}(AB)/2$ together with $\epsilon^{ij}\mbox{tr}(h_{\alpha_{ij}})
=\mbox{tr}(h_{\alpha_{12}})-\mbox{tr}(h_{\alpha_{12}}^{-1})=0$, a particular
property of $SU(2)$ (we did not need to use this, the result holds for 
general $G$).\\
Thus, any solution to the curvature constraint is a solution to the 
Hamiltonian constraint. However, as we have demonstrated, there are an 
infinite number of more solutions to the Hamiltonian constraint, in 
particular those which are normalizable with respect to $<.,.>_{Diff}$ 
and no solution to the curvature constraint has this property. Notice that 
the inner product on the space of solutions to the curvature constraint 
comes from a group averaging map, it is just given by (\cite{10})
$$
<\Psi_f,\Psi_g>_{Curv}:=\Psi_f(g)=\int_{{\cal M}} d\nu_0\overline{f} g\;.
$$
It is now tempting to view this result as the restriction to the special 
solutions of the curvature constraint of a more general inner product
appropriate for the Hamiltonian constraint. 

There seems to be an unsurmountable obstacle : the Hamiltonian constraint 
is not a self-adjoint operator on $\cal H$ and so group averaging as 
defined in $\cite{5}$ cannot be employed. Moreover, group-averaging 
really means to exponentiate the Hamiltonian constraint and that in turn 
implies that we know the motions it generates and thus we would have to 
completely solve the theory. Thus, it seems that we cannot 
define a map $\eta\; :\;\Phi\to\Phi'_{phys};\;f\to \eta f$. However, in the 
case that we have self-adjoint constraint operator, the group-average 
algorithm is nothing else than a sophisticated way to construct the 
projector onto the distributional kernel of the constraint operator
(this is explained in more detail in \cite{9}). We are therefore
lead to define the map $\eta$, in the case that we do not have a self-adjoint
constraint operator, as a certain (generalized) projector on the kernel of 
the constraint operator. As in \cite{10} we split the problem into two
parts and proceed as follows : \\
Given $f\in\Phi$ we have a group averaging map 
$\eta_{Diff}:\;\Phi\to\Phi'_{Diff}$ defined by $\eta_{Diff}(f):=[f]$
and an inner product defined by $<[f],[g]>_{Diff}:=[f](g):=<[f],g>
:=\int_\agb d\mu_0 \overline{[f]}g$. We define now 
$\Phi_{Ham}:=\Phi'_{Diff}$ and would like to define a map 
$\eta_{Ham}:\;\Phi_{Ham}\to\Phi'_{Ham}$. The space $\Phi'_{Ham}$ 
coincides with $\Phi'_{phys}$ when viewed as a space of distributions on
$\Phi$ via the map $\eta:=\eta_{Ham}\circ\eta_{Diff}$. It remains to
construct $\eta_{Ham}$. \\
As we have seen, the 
elements $[s]\in\Phi'_{phys},\;s\in S$ span $\Phi'_{phys}$. Moreover, by 
explicit construction (given for the 3+1 theory in \cite{2}) we can 
orthonormalize them with respect to $<.,.>_{Diff}$ thus exploiting that 
in our case all these $[s]$ are normalizable with respect to 
$<.,.>_{Diff}$. We 
obtain particular elements $\psi_\mu\in\Phi'_{Ham}\cap\Phi_{Ham}$ with 
the property that $<\psi_\mu,\psi_\nu>_{Diff}=\delta_{\mu,\nu}$. We are 
now ready to define the projector $\eta_{Ham}$ : given 
$\psi\in\Phi_{Ham}$ define
\be \label{52}
\eta_{Ham}\psi:=\sum_\mu \psi_\mu \psi_\mu(\psi):=\sum_\mu \psi_\mu 
<\psi_\mu,\psi>_{Diff}\;. \ee
Notice that even if not all of the $[s]$ would be normalizable with 
respect to $<.,.>_{Diff}$ then one could still take (\ref{52}) as the 
group average map, just the elements $\psi_\mu$ now form a basis in 
the generalized sense that they are mutually orthogonal in the 
sense of generalized eigenvectors (similar to usual 
momentum generalized eigenfunctions of ordinary quantum 
mechanics which are not really orhtonormal in the Hilbert space sense but 
only orthogonal in the sense of $\delta$ distributions).\\
The fact that the $\psi_\mu$ are normalizable with respect to $<.,.>_{Diff}$
displays $\eta_{Ham}$ as a projector on a genuine subspace of 
${\cal H}_{Diff}$.

Observe the dual role of the $\psi_\mu$ which we can view both as elements 
of $\Phi_{Ham}'$ and as elements of $\Phi_{Ham}=\Phi'_{Diff}\subset{\cal 
H}_{Diff}$. 
In particular, notice the peculiar identity $\eta_{Ham}\psi_\mu
=\psi_\mu$.\\
We now simply define an inner product on the elements $\eta_{Ham}\psi$ by 
\be \label {53}
<\eta_{Ham}\psi,\eta_{Ham}\psi'>_{Ham}:=(\eta_{Ham}\psi)(\psi')=
=\sum_\mu \overline{<\psi_\mu,\psi>_{Diff}}<\psi_\mu,\psi'>_{Diff}
\ee
for each $\psi,\psi'\in\Phi_{Ham}$. Expression (\ref{53}) is 
clearly a positive semi-definite sesquilinear form with the property that
the $\eta_{\psi_\mu}$ remain orthonormal. It is also independent of 
the orthonormal system $\psi_\mu$ (the label $\mu$ is ``nicely split" 
into a discrete piece and a continuous piece and $\psi_\mu$'s are 
orthonormal with respect to both pieces in the sense of Kronecker
$\delta$'s, see \cite{9} for details).\\ 
We now combine the two group average maps to obtain \be \label{54}
\eta:=\eta_{phys}:=\eta_{Ham}\circ\eta_{Diff}\;:\;\Phi:=\Phi_{phys}
\to \Phi'_{phys}:=\Phi'_{Ham},\;
f\to \eta_{Ham}[f]=\sum_\mu \psi_\mu \psi_\mu(f) 
\ee
and the physical inner product for the elements $\eta_{phys} f$ becomes
\ba \label{55}
<\eta_{phys} f,\eta_{phys} f>_{phys}&=&<\eta_{Ham}[f],\eta_{Ham}[g]>_{Ham}
\nonumber\\
&=&\sum_\mu \overline{<\psi_\mu,[f]>_{Diff}}<\psi_\mu,[g]>_{Diff}
\nonumber\\
&=&\sum_\mu \overline{\psi_\mu(f)}\psi_\mu(g)=(\eta_{phys}f)(g)
\ea
where in the second before the last equality $\psi_\mu$ is viewed as an 
element of $\Phi'$. \\
Notice that with this definition 
${\cal H}_{phys}\subset{\cal H}_{Diff}$. That this makes sense is shown
in \cite{9}. In other words, infinite linear combinations of elements 
of $\psi_\mu$ are allowed but only with suitably converging coefficients.\\
As we have already shown that $\Psi_f\not\in {\cal H}_{diff}$ it follows that
$\Psi_f\not\in{\cal H}_{phys}$, no solution to the curvature constraint
is normalizable with respect to $<.,.>_{Ham}$. 
Since what really determines the physical inner product is the 
Hamiltonian constraint, for instance via the group average approach, this
result was expected given the totally different algebraic structure of 
the two sets of constraints. In particular, the scalar product 
$<.,.>_{curv}$ is rather unnatural from the point of view of the
Hamiltonian constraint.\\
The reverse question, whether $||.||_{Ham}$ normalizable elements of 
$\Phi'_{Ham}$ have finite norm with respect to $||.||_{curv}$ cannot even 
be asked in general because a general element of $\Phi'_{Ham}$ cannot be 
written as $\Psi_f,\;f\in \Phi$.

We conclude that the sectors of the theory described by either of the 
inner products $<.,.>_{Curv}$ and $<.,.>_{phys}$
are mutually singular (that is, the underlying measures of 
the scalar products are singular). On the other hand, as far as the 
space of solutions to the constraint is concerned we find that all solutions
to the curvatur constraint are annihilated by the Hamiltonian constraint.
Moreover, if we choose 
$<.,.>_{Curv}$ as the inner product then we find complete agreement with
the results in \cite{0,1} although our set of constraints and our 
quantization approach was totally different from the outset. Since we copied 
step by step the quantization procedure of \cite{2} to maximal extent, we 
conclude that this procedure {\em does} lead to the correct answer in the   
present model which is a small but non-trivial check whether the 
proposal of \cite{2} is reliable or not.

\section{Conclusions}

The aim of the present paper was to check whether the method of quantizing
3+1 general relativity by the method proposed in \cite{2} is reliable
in the sense that when that procedure is applied to well-known models
we get the known the results. When applied to 2+1 Euclidean gravity we find 
complete agreement thus giving faith in those methods, the more, as 2+1 
Euclidean gravity is maximally similar to 3+1 Lorentzian gravity as far as 
the algebraic structure of the constraints and the gauge group are 
concerned (at least when we consider the Hamiltonian rather than the
curvature constraint). 

On the other hand, the quantum theory as obtained by our 
approach has a much bigger space of solutions to all constraints than the 
space as obtained by traditional approaches while the latter is properly 
included in our solution space. A natural question that arises is then 
what to do with those extra solutions and how to interprete them. 
In particular, there seems to be a clash between the number of classical 
and quantum degrees of freedom.

Now, a hint of how to interprete 
these solutions is that many of them are of the form $[s],\;s\in S$ and so
$s$ is a cylindrical function. Therefore the volume operator $\hat{V}(B)$ 
vanishes on $s$ for almost every $B$. This suggests that $[s]$ is a 
spurious solution because if we want the classical theory defined by
curvature and Hamiltonian and diffeomorphism constraint to be equivalent 
(recall that we had to impose $\det(q)>0$ classically)
then we must really ask that the volume operator $\hat{V}(B)$ is strictly 
positive for any $B$, before taking the diffeomorphism constraint into 
account. 
We conclude that no cylindrical $s$ should give rise to an element of 
$\Phi'_{phys}$ via $s\to [s]$ (which can be achieved by superposition
of an infinite number of the $[s]$ or by considering infinite graphs) 
which would presumably remove the clsh between numbers of degrees of 
freedom alluded to above.
This latter observation gives rise to the speculation that also many of 
the solutions found in \cite{2} for the 3+1 theory should be spurious 
because in the classical
theory we have to impose the anholonomic condition $\det(q)>0$ as well.
On the other hand it may be desirable to allow for degenerate solutions 
at the quantum level because by passing through singularities of the 
metric one can describe changes in the topology of the hypersurface 
$\Sigma$. Therefore one may expect that some of the solutions actually 
carry topological information and, moreover, that although we have 
started from a fixed topology in the classical theory we end up describing 
all topologies at the quantum level\footnote{This speculation on the 
conceivably topological meaning of the solutions is due to Abhay
Ashtekar.}. 
The complete answer to this puzzle is left to future investigations.\\ 
\\ 
\\
\\
{\large Acknowledgements}\\
\\
This project was suggested by Abhay Ashtekar during the wonderful 
ESI workshop in Vienna, July-August 1997. Many thanks to the ESI and to
the organizers of the workshop, Peter Aichelburg and Abhay Ashtekar, 
for providing this excellent research environment.
This research project was supported in part by the ESI and by DOE-Grant
DE-FG02-94ER25228 to Harvard University.

\begin{appendix}

\section{Spectral analysis of the two-dimensional volume operator}

Since the gauge group is still $SU(2)$ we may copy the results from \cite{7}
to compute the full spectrum of the two-dimensional volume operator. In 
particular it follows immediately that this operator is essentially 
self-adjoint, positive semi-definite and that its spectrum is entirely 
discrete. This holds on either gauge invariant or non-gauge invariant
functions.\\
In this appendix we restrict ourselves to the part of the spectrum coming 
from graphs with vertices of valence not larger than three, that is, we 
display the eigenvalues of the operator $\hat{V}_v$
of (\ref{32}) restricted to vertices $v$ of valence $n=2,3$ on gauge 
invariant functions. Indeed, as the volume operator cannot change the 
graph or the colouring of the edges of the graph with spin quantum 
numbers of a spin-network state it follows that it can change at most its 
vertex contractors. However, given the spins of the edges incident at $v$,
the space of vertex contractors is one-dimensional for $n=2,3$ by 
elementary Clebsh-Gordon theory. Therefore spin-network states all of 
whose vertices have at most valence three must be eigenvectors of the 
volume operator (in any dimension). Notice also that all the $\hat{V}_v$ for 
different $v$'s are mutually commuting. In three dimensions 
these spin-network states are in the kernel of the volume operator, in two
dimensions none of them is annihilated as we will show (as long as the 
tangents of the edges at $v$ span a plane).

For $n=3$ there are only two generic non-trivial situations : Either 
(Case A) no two of $e_1,e_2,e_3$ have co-linear tangents at $v$ or
(Case B) two of them, say $e_1,e_2$ have co-linear tangents at $v$ but not
$e_1,e_3$ or $e_2,e_3$. Here $e_1,e_2,e_3$ are the three edges incident 
at $v$ which are coloured with spins $j_1,j_2,j_3\in\{j_1+j_2,j_1+j_2-1,..,
|j_1-j_2|\}$ respectively. We can get the 
eigenvalue for the case $n=2$ by taking the result for $n=3$ and setting 
for instance $j_3=0,j_1=j_2=j\not=0$.\\
In the calculations that follow we will use the following notation :
\ba \label{a1}
\hat{q}_v&:=&[\frac{4}{\hbar^2}\hat{V}_v]^2=\hat{E}_v^i\hat{E}_v^i
\nonumber\\
\hat{E}_v^i&:=&\frac{1}{2}\sum_{1\le I,J\le 3} \mbox{sgn}(e_I,e_J)
X^i_{IJ}\mbox{ where }X^i_{IJ}:=\epsilon_{ijk}X^j_I X^k_J\nonumber\\
X^i_I&:=& X^i_{e_I},\;\vec{X}_I:=(X^i_I),\; X_{IJ}=X^i_I X^i_J,\; 
\Delta_I:=X_{II} \ea
and it is implied that $I,J,K\in\{1,2,3\}$ are mutually different so that
$[X^i_I,X^j_J]=0$. As the notation suggests, $\Delta_I$ is the Laplacian
on $SU(2)$ with spectrum $-j(j+1),\;2j\ge 0$ integral.
Notice that $X^i_{IJ}=-X^i_{JI}$ so that 
$$
\hat{E}_v^i=\sum_{1\le I<J\le 3} \mbox{sgn}(e_I,e_J) X^i_{IJ}\;.
$$
As in the main text we will use generators of $su(2)$ with structure 
constants $+\epsilon_{ijk}$ which implies that $[X^i_I,X^j_I]=-\epsilon_{ijk}
X^k_I$ and so $\epsilon_{ijk}X^i_I X^j_I=-X^k_I$ (the minus sign comes 
from the {\em right} rather then left invariance).\\
There are some identities among these quantities that we are going to use.
The first one is the familiar spin recoupling identity 
\be \label{a2}
2X_{IJ}=[\vec{X}_I+\vec{X}_J]^2-\Delta_I-\Delta_J=
\Delta_K-\Delta_I-\Delta_J
\ee
where in the second equality we have used the fact that 
$\vec{X}_1+\vec{X}_2+\vec{X}_3=0$, that is, the total angular momentum 
operator vanishes of on gauge invariant functions. Then if $f$ is gauge 
invariant 
$$
[\vec{X}_I+\vec{X}_J]^2 f=-[\vec{X}_I+\vec{X}_J]\vec{X}_K f=
-\vec{X}_K[\vec{X}_I+\vec{X}_J]f=[\vec{X}_K]^2 f=\Delta_K f
$$
and of course the $\Delta_I$ commute with every $X^i_I$. The next identity
is, using basic $\epsilon_{ijk}$ arithmetic
\be \label{a3}
\vec{X}_{IJ}^2=X^i_I X^j_J(X_I^i X^j_J-X^j_I X^i_J)=\Delta_I\Delta_J
-X^i_I([X^j_J,X^i_J]+X^i_J X^j_J)X^j_I=\Delta_I\Delta_J
+X_{IJ}-X_{IJ}^2
\ee 
and by very similar arguments
\be \label{a4}
\vec{X}_{IJ}\vec{X}_{JK}=-\Delta_J X_{IK}+X_{IJ} X_{JK}
+\epsilon_{ijk}X^i_I X^j_K X^k_J\;.
\ee
The last term in (\ref{a4}) is essentially the basic operator from which 
the tree-dimensional volume operator is built and which vanishes in the 
three-valent case on gauge invariant functions. Indeed, replacing, say
$\vec{X}_J=-\vec{X}_I-\vec{X}_K$ and using the $su(2)$ algebra for the 
$\vec{X}^i_I$ we see that that term vanishes. 

Remarkably, upon substituting
for $X_{IJ}$ according to (\ref{a2}) we find
\be \label{a5}
\vec{X}_{IJ}\vec{X}_{JK}=
\frac{1}{4}[2(\Delta_I\Delta_J+\Delta_J\Delta_K+\Delta_K\Delta_I)
-(\Delta_I^2+\Delta_J^2+\Delta_K^2)]
\ee
which is {\em independent} of the choice of the pairs $(IJ),(JK)$.

We have now all tools available to finish the calculation. 
We will treat cases A, B separately.
\begin{itemize}
\item[A)]
We may label edges without loss of generality such that 
$\mbox{sgn}(e_1,e_2)=\mbox{sgn}(e_2,e_3)=\mbox{sgn}(e_3,e_1)=1$, that is,
we cross $e_1,e_2,e_3$ in this sequence as we encircle $v$ counter-clockwise.
Then $\vec{\hat{E}}_v=\vec{X}_{12}+\vec{X}_{23}+\vec{X}_{31}$. We just need 
to use (\ref{a2})-(\ref{a5}) and to be careful with the order of $I,J$ in
$\vec{X}_{IJ}$ to find after tedious algebra that 
$\hat{q}_v=[\vec{\hat{E}}_v]^2$ is just given by
\ba \label{a6}
\hat{q}_v&=&\frac{9}{4}
[2(\Delta_1\Delta_2+\Delta_2\Delta_3+\Delta_3\Delta_1)
-(\Delta_1^2+\Delta_2^2+\Delta_3^2)]\nonumber\\
&-& \frac{1}{2}(\Delta_1+\Delta_2+\Delta_3)\;.
\ea
Thus, the eigenvalue is obtained by replacing $\Delta_I$ by $-j_I(j_I+1)$.
Expression (\ref{6}) looks worrysome : is the eigenvalue going to be 
non-negative ? A moment of reflection reveals that it is even strictly 
positive
unless $j_1=j_2=j_3=0$ in which case it vanishes : It will be sufficient to 
show that the operator 
in the first line of (\ref{a6}) has non-negative eigenvalue. We just need 
to remember that $j_1,j_2,j_3$ are not arbitrary. We may assume without loss
of generality that $j_2\ge j_1$ such that $j_3\in\{j_1+j_2,j_1+j_2-1,..,
j_2-j_1\}$. We have 
\be \label{a7}
f(\Delta_3):=2(\Delta_1\Delta_2+\Delta_2\Delta_3+\Delta_3\Delta_1)
-(\Delta_1^2+\Delta_2^2+\Delta_3^2)=4\Delta_1\Delta_2
-(\Delta_3-\Delta_1-\Delta_2)^2
\ee
which takes, in terms of eigenvalues, its lowest value at maximum value
of the function $|\Delta_3-\Delta_1-\Delta_2|$. Given arbitrary $j_1\le j_2$,
since $-\Delta_3$ is a strictly increasing function of $j_3$, we find that
the extrema of that function are found for the extremal values $j_3=j_2\pm 
j_1$ and are given by $|2j_1 j_2|$ and $|-2 j_1(j_2+1)|$ respectively.
Then (\ref{a7}) reveals that $f(\Delta_3)\ge 
4j_1(j_2+1)(j_2-j_1)\ge 0$ because $j_2\ge j_1$.

In case that we consider a two-valent vertex, we may just set $\Delta_3=0,\;
\Delta_1=\Delta_2=\Delta$ and find the extremely simple result
\be \label{a8}
\hat{q}_v=-\Delta\;.
\ee
\item[B)]
We may, without loss of generality, label edges such that $e_1,e_2$ have 
co-linear tangents at $v$ (that is, $\mbox{sgn}(e_1,e_2)=0$) and such that 
$\mbox{sgn}(e_1,e_3)=\mbox{sgn}(e_3,e_2)=1$. Then 
$\vec{\hat{E}}_v=\vec{X}_{13}+\vec{X}_{32}$. The same algebraic 
manipulations show that we get now for $\hat{q}_v=[\vec{\hat{E}}_v]^2$
the expression
\be \label{a9}
\hat{q}_v=
[2(\Delta_1\Delta_2+\Delta_2\Delta_3+\Delta_3\Delta_1)
-(\Delta_1^2+\Delta_2^2+\Delta_3^2)]-\Delta_3
\ee
which is positive unless, of course, $\Delta_3=0$ in which case it vanishes.
\end{itemize}
We conclude that the two-dimensional volume operator has a much smaller
kernel than the three-dimensional one, in particular, two and three-valent
vertices, whether gauge invariant or not, do not contribute to the kernel.

\end{appendix}


\begin{thebibliography}{99}

\parskip -5pt


\bibitem{0} S. Carlip, ``Lectures in (2+1)-Dimensional Gravity", Preprint
UCD-95-6, grqc/9503024\\
S. Carlip, ``The Statistical Mechanics of the Three-Dimensional Euclidean 
Black Hole", Preprint UCD-96-13, gr-qc/9606043\\
V. Moncrief, J. Math. Phys. {\bf 30} (1989) 2907

\bibitem{1} E. Witten, Nucl. Phys. {\bf B311} (1988) 46

\bibitem{2} T. Thiemann, Phys. Lett. B {\bf 380} (1996) 257-264\\
T. Thiemann, ``Quantum Spin Dynamics (QSD)", Harvard 
University Preprint HUTMP-96/B-359, gr-qc/9606089\\
T. Thiemann, ``Quantum Spin Dynamics (QSD) II : The Kernel of the 
Wheeler-DeWitt Constraint Operator", 
Harvard University Preprint HUTMP-96/B-352, gr-qc/9606090

\bibitem{3} A.\ Ashtekar, V.\ Husain, C.\ Rovelli, J.\ Samuel, L.\ Smolin,
            Class. Quantum Grav.\ {\bf 6},  L183 (1989).

\bibitem{4} F. Barbero, M. Varadarajan, ``The Phase Space off 2+1 Dimensional
Gravity in the Ashtekar Formulation, Nucl. Phys. {\bf B415} (1994) 515

\bibitem{5} A. Ashtekar, J. Lewandowski, D. Marolf, J. Mour\~ao, T.
Thiemann, ``Quantization for diffeomorphism invariant theories 
of connections with local degrees of freedom", Journ. Math. Phys.
{\bf 36} (1995) 519-551

\bibitem{5a} A. Ashtekar and C.J. Isham,
Class. \& Quan. Grav. {\bf 9}, 1433 (1992)\\
A. Ashtekar and J. Lewandowski, ``Representation
theory of analytic holonomy $C^\star$ algebras'', in {\it Knots and
quantum gravity}, J. Baez (ed), (Oxford University Press, Oxford 1994)\\
A. Ashtekar and J. Lewandowski, ``Differential
geometry on the space of connections via graphs and projective
limits'', Journ. Geo. Physics {\bf 17} (1995) 191\\
A. Ashtekar and J. Lewandowski, J. Math. Phys. {\bf 36}, 2170
(1995). \\
D. Marolf and J. M. Mour\~ao, ``On the support of the
Ashtekar-Lewandowski measure'',  Commun. Math. Phys. {\bf 170} (1995)
583-606\\
A. Ashtekar, J. Lewandowski, D. Marolf, J. Mour\~ao
and T. Thiemann, ``A manifestly gauge invariant approach to quantum
theories of gauge fields'', in {\it Geometry of constrained dynamical
systems}, J. Charap (ed) (Cambridge University Press, Cambridge,
1994); ``Constructive quantum gauge field theory in two space-time
dimensions'' (CGPG preprint).

\bibitem{6} C.\ Rovelli, L.\ Smolin, ``Discreteness of volume and 
area in quantum gravity'' Nucl. Phys. B {\bf 442} (1995) 593, Erratum :
Nucl. Phys. B {\bf 456} (1995) 734\\
A. Ashtekar, J. Lewandowski, ``Quantum Geometry III : Volume Operators",
(in preparation)\\
J. Lewandowski, ``Volume and Quantizations", Class. Quantum Grav. {\bf 14}
(1997) 71-76\\
R. De Pietri, C. Rovelli, ``Geometry eigenvalues and scalar product from 
recoupling 
theory in loop quantum theory", Reprint UPRF-96-444, gr-qc/9602023

\bibitem{7} T. Thiemann, ``Complete formula for the matrix elements of the 
volume operator in canonical quantum gravity", Harvard University Preprint 
HUTMP-96/B-353 gr-qc/9606091

\bibitem{TTMat} T. Thiemann, ``QSD V : Quantum Gravity as the Natural 
Regulator of Matter Quantum Field Theories", Harvard University Preprint
HUTMP-96/B-357

\bibitem{8} T. Thiemann, ``A length operator for canonical quantum gravity",
Harvard University Preprint HUTMP-96/B-354, gr-qc/9606092

\bibitem{9} T. Thiemann, ``QSD III : Quantum Constraint Algebra and 
Physical Scalar Product in Quantum General Relativity", 
Harvard University Preprint HUTMP-97/B-363

\bibitem{10} D. Marolf, J. Mour\~ao, T. Thiemann, ``The status of 
Diffeomorphism Superselection in Euclidean 2+1 Gravity", HUTMP-97/B-360,
gr-qc/9701068




\end{thebibliography}
\end{document}